\documentclass[journal=jacsat,manuscript=article]{achemso}

\usepackage[version=3]{mhchem} 
\usepackage[english]{babel}
\usepackage{amsfonts}
\usepackage{enumitem,booktabs,cfr-lm}
\usepackage[referable]{threeparttablex}
\renewlist{tablenotes}{enumerate}{1}
\makeatletter
\setlist[tablenotes]{label=\tnote{\alph*},ref=\alph*,itemsep=\z@,topsep=\z@skip,partopsep=\z@skip,parsep=\z@,itemindent=\z@,labelindent=\tabcolsep,labelsep=.2em,leftmargin=*,align=left,before={\footnotesize}}
\makeatother
\usepackage{verbatim}
\usepackage{amsmath}
\usepackage{amsthm}
\usepackage{amssymb}
\usepackage{graphicx}
\usepackage{subfigure}
\usepackage{cancel}
\usepackage{physics}
\usepackage{siunitx}
\usepackage[usenames,dvipsnames]{color}
\usepackage{color,linegoal}
\usepackage[procnames]{listings}
\SectionNumbersOn

\definecolor{myc}{RGB}{0,0,0}

\usepackage{natbib}

\author{Alessandro~Carbonaro}
\affiliation{Laboratoire Charles Coulomb (L2C), UMR 5221 CNRS-Universit\'{e} de Montpellier,
Montpellier, France}
\author{Luca~Cipelletti}
\affiliation{Laboratoire Charles Coulomb (L2C), UMR 5221 CNRS-Universit\'{e} de Montpellier,
Montpellier, France}
\author{Domenico~Truzzolillo}
\email{domenico.truzzolillo@umontpellier.fr}
\affiliation{Laboratoire Charles Coulomb (L2C), UMR 5221 CNRS-Universit\'{e} de Montpellier,
Montpellier, France}

\title{Spinning drop dynamics in miscible and immiscible environments}

\begin{document}
\begin{abstract}
We report on the extensional dynamics of spinning drops in miscible and immiscible background fluids following a rotation speed jump.
Two radically different behaviours are observed. Drops in immiscible environments relax exponentially to their equilibrium shape, with a relaxation time that does not depend on the centrifugal forcing. We find an excellent quantitative agreement with the relaxation time predicted for quasi-spherical drops by Stone and Bush (Q. Appl. Math. 54, 551 (1996)), while other models proposed in the literature fail to capture our data.
By contrast, drops immersed in a miscible background fluid do not relax to a steady shape: they elongate indefinitely, their length following a power-law $l(t)\sim t^{\frac{2}{5}}$ in very good agreement with the dynamics predicted by Lister and Stone (J. Fluid Mech. 317, 275 (1996)) for inviscid drops. 
Our results strongly suggest that low compositional gradients in miscible fluids do not give rise to an effective interfacial tension measurable by spinning drop tensiometry.
\end{abstract}

\section{Introduction}

Over the last century, liquid droplets have attracted the attention of physicists \cite{rayleigh_equilibrium_1914}, since they are fairly simple to study and control under different conditions, and because they represent a model system to understand the physics of phenomena occurring on different scales, ranging from laboratory \cite{janiaud_spinning_2000,hill_nonaxisymmetric_2008} to astrophysical ones \cite{chandrasekhar_hydrodynamic_1961}. A droplet, for example, can spread out or ball up and spin depending on the interaction with the surface it lies on \cite{li_spontaneous_2019} and its shape can be drastically modified by external fields \cite{filali_deformation_2018} or when it is set in motion \cite{kekesi_drop_2016, holgate_shapes_2018}. In particular, rotating droplets are among the most studied cases of liquids deformed by an external field as they are relevant in many situations: rotation plays a pivotal role on the structure and evolution of large-scale flows taking place in oceans, atmosphere and in the very body of planets and stars \cite{zhang_theory_2017, chandrasekhar_hydrodynamic_1961}.

Freely-suspended droplets rotating at high speed tend to deform due to centrifugal forcing. Droplets change shape following a minimum energy principle, taking the lowest energy state for a given rotational frequency. This results into a series of non-trivial equilibrium shapes, strongly affected by surface tension \cite{brown_shape_1980,hill_nonaxisymmetric_2008}.
The equilibrium shape of a weightless spinning droplet under the action of capillary forces was first discussed by Lord Rayleigh (1914) \cite{rayleigh_equilibrium_1914}. He found a solution in which the bubble is a surface of revolution which meets the axis of rotation. If the angular speed is zero, the bubble has a spherical form, while under the influence of rotation the sphere elongates along the rotation axis, and the oblateness increases upon increasing the angular velocity. However, more complicated shapes are obtained above a critical velocity that depends on interfacial tension and drop size, as shown by Hill et al. \cite{hill_nonaxisymmetric_2008} for magnetically levitated water droplets. Magnetic levitation is indeed one appealing way to isolate and suspend single drops, however it is experimentally challenging, requiring the use of strong magnetic fields, especially for molecular fluids that display weak diamagnetic properties.

An alternative experimental configuration is that of a drop of liquid immersed in a denser background fluid, both confined in a horizontal cylindrical capillary. When the capillary is set in rotation, centrifugal forces confine the drop on the axis, avoiding any contact with the capillary wall. Furthermore, the shape of the drop is dictated by the balance between the centrifugal force and interfacial tension, allowing the latter to be measured. This is the principle of spinning drop tensiometry, originally introduced by Vonnegut \cite{vonnegut_rotating_1942} to measure interfacial tensions between immiscible fluids and further developed by Princen et al. \cite{princen_measurement_1967}, Torza \cite{torza_rotatingbubble_1975}, and Cayias et al. \cite{cayas_acs_nodate}. Thanks to their work, spinning drop tensiometry (SDT) has become a versatile method, particularly well suited for measuring ultralow interfacial tensions (down to 10 $\mu$N/m). Such low tensions can occur, for example, in water-hydrocarbon-surfactant systems and are of considerable scientific interest as well as of great importance for industrial applications.

While the majority of works on spinning drops tensiometry focussed on  the equilibrium shape of the drops, several authors have characterized their relaxation after a jump in rotational speed, when suspended in an immiscible environment \cite{patterson_measurement_1971,hsu_rheological_1975,joseph_spinning_1992,joseph_upper_1992,hu_evolution_1994}. However, a general consensus on the mechanisms driving drop relaxation is still missing and different models have been proposed to capture the experimental behaviour of Newtonian drops for both small and large deformations \cite{joseph_spinning_1992,joseph_upper_1992,hu_evolution_1994,stone_stone_1996}.
On the experimental side, Joseph et al. \cite{joseph_spinning_1992} performed relaxation tests using polydimethyl siloxane ($\eta=$10$^5$ mPa s) drops in glycerol, a system similar to that investigated in the present work, using a spinning drop tensiometer described in Ref. \cite{than_measurement_1988} at only one angular speed ($\omega=$ 3208 rpm). They found that the drop relaxes exponentially to its equilibrium shape. The same authors \cite{joseph_spinning_1992} reported a similar exponential relaxation for drops of polystyrene in poly(methyl methacrylate) (PMMA), drops of standard oil in water-glycerol mixtures \cite{hsu_rheological_1975} and drops of low-density polyethylene in PMMA \cite{verdier_topics_1990}. In all cases, the value of the relaxation constant departed from that predicted theoretically by the same authors for largely deformed drops (see below Eq. \ref{Josephrelax}). To summarize, several experiments suggest that largely deformed spinning drops in immiscible fluids relax exponentially towards an equilibrium shape after a jump of centrifugal forcing. However, a quantitative description of the relaxation time and a clear understanding of the relevant parameters that determine its value are still missing.

Research on drops in a miscible background fluid is less advanced. A full experimental characterization of drops showing pure extensional dynamics is missing: indeed, in most cases miscible drops undergo also a radial deformation due to secondary flows that set in a spinning capillary~\cite{manning_interfacial_1977} and this greatly complicates both measurements and modelling. P. Petitjeans \cite{petitjeans_petitjean_1996} was the first to study fluids miscible in all proportions with SDT. In a series of measurements involving drops of water immersed in water-glycerol mixtures, he observed the stabilization of the drop radius after about 100 seconds for mixtures containing more than 40 \% glycerin mass fraction. By measuring the drop radius and using Vonnegut's formula (see Section \ref{theory} below), he inferred the effective (transient) interfacial tension between the fluids. Similar measurements were performed by J. Pojman and co-workers \cite{pojman_evidence_2006,zoltowski_evidence_2007}, who investigated the existence of an effective interfacial tension (EIT) between miscible fluids, using i) mixtures of Isobutyric acid and water close to their critical point and ii) dodecyl acrylate drops in poly(dodecyl acrylate). They concluded that capillary forces are at work at the boundary between these fluids and quantified the EIT, again using Vonnegut's theory.

An important assumption in Refs.~\cite{petitjeans_petitjean_1996,pojman_evidence_2006,zoltowski_evidence_2007} is the fact that the drop reaches a quasi-equilibrium state or at least a steady state under rotation. Unfortunately, in these works
the temporal evolution of the drop length is not reported: monitoring it would allow one to unambiguously ascertain the existence of such a steady state. Moreover, Pojman and co-workers observed the formation of dog-bone-shaped drops~\cite{zoltowski_evidence_2007}, suggesting that secondary viscous flows may perturb the system. Such effect has been first pointed out by Manning et al.~\cite{manning_interfacial_1977}, who discussed the role of secondary viscous flows in spinning drop tensiometry, underlining that they may lead to significant deviations with respect to the ideal case of Vonnegut's theory even for immiscible fluids, provided that the interfacial tensions is sufficiently low, $\Gamma \lesssim 10~\mu$N/m.

In this work, we circumvent these difficulties and systematically investigate the temporal evolution of the drop length in SDT experiments that probe both immiscible and miscible systems, aiming in particular at detecting the existence of transient capillary forces at the interface between miscible fluids. Using a custom imaging setup, we follow the evolution of very elongated drops, assessing unambiguously whether or not a stationary state is reached. By focussing on the drop length, we furthermore avoid the complications inherent to the measurement of the drop radius, stemming from the curvature of the cylindrical capillary-air dioptre. We compare the results of our experiments to  existing theories for the stretching behaviour of drops, for both the immiscible and miscible cases. We find that immiscible drops relax to their equilibrium shape exponentially, with a characteristic relaxation time $\tau$ independent of both the magnitude of the jump in $\omega$ and the equilibrium length of the drop. Our experiments therefore support the notion that for a given pair of fluids $\tau$ is an intrinsic property of the system, in excellent agreement with the prediction for small deformations of quasi-spherical drops by Stone and Bush~\cite{stone_stone_1996}. The scenario for miscible fluids is strikingly different. We find that the drop shape never reaches a stationary state. Instead, for drops with a low concentration gradient with respect to the background fluid, the drop length increases indefinitely following asymptotically a power law, $l(t)\sim t^{\frac{2}{5}}$. The drop retains a cylindrical shape with spherical endcaps, indicating that in the experiments reported here radial deformations due to secondary flows are absent or negligible. The elongation behaviour of our miscible drops is very well captured by a model originally proposed by Lister and Stone~\cite{lister_time-dependent_1996} for immiscible fluids, in the limit of vanishing surface tension and vanishing viscosity of the drop fluid (``bubble-like'' dynamics).

The rest of the paper is organized as follows. Section \ref{theory} briefly recalls the theoretical background and the existent predictions for the elongational dynamics of spinning drops. In section \ref{matmeth} we present the materials employed and the experimental setup. Sections \ref{immiscible} and \ref{miscible} present the results on the elongational dynamics for drops in an immiscible and miscible background fluid, respectively. Section \ref{conclusions} concludes the paper.

\section{Theoretical background}\label{theory}
Spinning drop tensiometry assumes the gyrostatic equilibrium of the drop, i.e. a state of uniform rotation in which every fluid element in a spinning rigid container is at rest with respect to the container wall. Consider a small drop of fluid A placed in a more dense fluid B contained in a cylindrical capillary. When the capillary is spun around its axis, the drop elongates axially and takes an ellipsoidal shape. At gyrostatic equilibrium, the normal component of the interfacial tension balances the hydrostatic pressure difference across the interface. If the drop length $l$ (measured along the direction of the capillary axis) exceeds four times its equatorial diameter $2r$ (measured perpendicularly to the capillary axis), the magnitude $\Gamma$ of the interfacial tension can be calculated very simply from the drop radius $r$, the density difference $ \Delta \rho= \rho_B - \rho_A$, and the angular velocity $\omega$, via the well known expression
\begin{equation}\label{Vonnegut}
\Gamma=\frac{\Delta\rho\omega^2r^3}{4}
\end{equation}
first derived by Vonnegut \cite{vonnegut_rotating_1942}. Note that other methods of determining the interfacial tension based on the shape of a drop require a measurement of the two- or even three-dimensional profile of the drop and involve complicated calculations, e.g. in the pendant drop method \cite{saad_design_2011}. By contrast, the simplicity of Vonnegut's expression is one of the main advantages of the spinning-drop method.

Eq. \ref{Vonnegut} holds at equilibrium, once a steady state has bean reached. The time evolution of the drop upon a step change of $\omega$ has also been investigated.
Theoretical models \cite{stone_stone_1996,hu_evolution_1994,joseph_spinning_1992} and experiments \cite{patterson_measurement_1971,hsu_rheological_1975,joseph_upper_1992}, both for immiscible fluids, suggest that drops relax exponentially towards their equilibrium state. Nevertheless, models developed for nearly spherical~\cite{stone_stone_1996} or very elongated~\cite{hu_evolution_1994} drops predict qualitatively different expressions for the typical relaxation time $\tau$. Stone and Bush~\cite{stone_stone_1996} derived an exact solution for the fluid motion and time-dependent drop shape, provided that the following conditions are fulfilled: the drop remains nearly spherical, wall effects can be neglected, and fluid flows both inside and outside the drop are dominated by viscous effects. The last assumption allows the drop dynamics to be described in terms of centrifugally-forced Stokes flows. In this case, the relaxation time after a rotation speed jump $\Delta \omega=|\omega_f-\omega_i|$ of a drop with viscosity $\eta_d$ immersed in a surrounding liquid with viscosity $\eta_e$ reads:
\begin{equation}\label{Stonerelax}
\tau=\frac{\eta_e a}{\Gamma}f(\lambda)\,,
\end{equation}
where $a$ is the radius of the drop at rest, $\lambda=\eta_d/\eta_e$ the viscosity contrast, and $f(\lambda)=\frac{(3+2\lambda)(16+19\lambda)}{40(1+\lambda)}$ a dimensionless prefactor. The relaxation time described by Eq.~\ref{Stonerelax} turns out to be identical to that obtained for drops in an extensional flow \cite{rallison_deformation_1984,stone_stone_1996}. Interestingly, according to Eq.~\ref{Stonerelax}, $\tau$ does not depend neither on the angular speed $\omega$ nor on its change $\Delta \omega$.

The theory of Ref.~\cite{stone_stone_1996} is limited to nearly spherical shapes. Hu et Jospeh have investigated the opposite limit of very elongated drops~\cite{hu_evolution_1994}, proposing a semi-empirical expression to fit the results of numerical simulations and experiments. The relaxation time may be cast in the form
\begin{equation}\label{Josephrelax}
\tau=\frac{\eta_d r_{eq}}{\Gamma}f_J(\lambda,l_{eq},r_{eq},R_c) \,,
\end{equation}
where $f_J(\lambda,l_{eq},r_{eq},R_c)$ is a function of the viscosity contrast, the inner radius of the cylindrical container, $R_c$, and the radius, $r_{eq}$, and length, $l_{eq}$, of the drop once equilibrium is reached at the final rotation speed $\omega_f$. This expression was derived for elongated drops, contrary to the theory of Ref.~\cite{stone_stone_1996}. However, both theories deal with the case where the change of the drop shape results from a small speed jump. Equation~\ref{Josephrelax} is qualitatively different from Eq.~\ref{Stonerelax}, since it predicts a dependence of $\tau$ on the equilibrium shape of the drop via $r_{eq}$ and $l_{eq}$, and hence on $\omega_f$. Hu and Joseph noticed that their theoretical expression for $\tau$ typically underestimates the time observed in experiments~\cite{hu_evolution_1994,joseph_spinning_1992}. They argued that the detailed form of the flow around the endcaps of the drop may slow down its relaxation and account for this discrepancy. 

So far, we briefly reviewed the theoretical background for the case of immiscible fluids displaying a non-negligible surface tension, for which spinning drop tensiometry was originally conceived. However, SDT has been also used, albeit much less frequently, to investigate the drop shape in miscible environments, aiming at measuring capillary forces between miscible fluids \cite{petitjeans_petitjean_1996,pojman_evidence_2006,zoltowski_evidence_2007}. Unfortunately, there is no theoretical prediction available so far for this case. Capillary phenomena are expected to be weak, if not negligible, in miscible fluids. It is therefore interesting to briefly recall available results on the behaviour of immiscible drops in the limit of vanishing interfacial tension. We focus mostly on the ``bubble'' limit, i.e. for vanishing viscosity ratio $\lambda$ between the drop and the surrounding fluid, and for intermediate ratios, as these cases will turn out to be of interest in our experiments.

In the cases of vanishing or intermediate viscosity ratios, Lister and Stone \cite{lister_time-dependent_1996} have obtained asymptotic scaling laws for spinning drops in an unbounded geometry.
The drops are assumed to be long and slender, with (time-dependent) equatorial radius $a_0$ much smaller than the length $l$. At all times, the interface between the two fluids is supposed infinitely sharp and stable, and surface tension is supposed to be negligible.
The rate of elongation and thinning is estimated by imposing a balance between the centrifugal pressure
$\frac{1}{2}\Delta \rho \omega^2 a_0^2$ times the area $\mathcal{O}(\pi a_0^2)$ over which it acts, and the dominant resisting
viscous stress times the area over which it acts. The viscous stress is found to depend on both the viscosity ratio $\lambda$ and the aspect ratio $l/2a_0$. We anticipate here that for the drop-background fluid system that we will consider here after, namely a drop containing 5\% wt H\textsubscript{2}O and 95\% wt glycerol suspended in pure glycerol, $\lambda=0.537$ and $1\lesssim l/2a_0 \lesssim 50$.

As it will turn out to be relevant for our drops, we first discuss the low viscosity ratio limit ($\lambda \ll 2a_0/l \ll 1$), for which the motion of the internal fluid as the drop extends generates less viscous dissipation than the deformation of the external fluid. Because of the centrifugal pressure, the shape of very elongated drops is expected to be close to that of a cylinder with domed endcaps. For small $\lambda$, the primary resistance to deformation is due to the displacement of the background fluid as the drop elongates.
A force balance in the neighbourhood of the endcap, which has radius $\mathcal{O}(a_0)$, surface
area $\mathcal{O}(2\pi a_0^2)$ and moves at velocity $U=\frac{1}{2}dl/dt$, yields $2\pi a_0^2 \eta_e U/a_0\simeq \frac{1}{2}\Delta\rho\omega^2a_0^2\pi a_0^2$.
Using volume conservation in the approximate form $V\simeq 2\pi a_0^2 l$ and estimating the velocity as $U\simeq l/2t$
yields the time-dependent length

\begin{equation}\label{lenghtmisc1}
    l(t)\simeq \left(\frac{\Delta\rho\omega^2V^{3/2}}{\eta_e}\right)^{\frac{2}{5}} t^{\frac{2}{5}}\,,
\end{equation}
where for clarity we have dropped multiplicative numerical constants.

For more viscous drops, the shear stresses generated by the motion of the internal fluid along the rotation axis towards the drop ends become relevant and along the (half) length of the drop a pressure gradient $\Delta\rho\omega^2a_0^2/l$ is established. The shear within the drop occurs on a lengthscale $\mathcal{O}(a_0)$, whereas the external shear occurs on a somewhat larger scale $\mathcal{O}(a_0\ln(l/a_0))$, as shown in Refs. \cite{batchelor_slender-body_1970,cox_motion_1970} by analysing the axial motion of a slender body with radius $a_0$ and length $l$.
The relative magnitudes of $\lambda$ and $1/\ln(l/2a_0)$ controls which shear term contributes most.
For the case $\lambda \ll 1/\ln(l/2a_0)$, the internal shear is dominant and is $\mathcal{O}(\eta_d U/a_0)$, while the total resistance along the whole drop length is $\mathcal{O}(\pi l \eta_d U)$. The rate of extension is then \cite{lister_time-dependent_1996}
\begin{equation}\label{lenghtmisc2}
    l(t)\simeq \left(\frac{\Delta\rho\omega^2V^2}{\eta_d}\right)^{\frac{1}{4}} t^{\frac{1}{4}}\,.
\end{equation}
In the opposite case, $\lambda\gg 1/\ln(l/2a_0)$, similar estimates lead to \cite{lister_time-dependent_1996}
\begin{equation}\label{lenghtmisc2b}
    l(t)\simeq \left(\frac{\Delta\rho\omega^2V^2}{\eta_e}\right)^{\frac{1}{4}} \left(t\ln\frac{t}{\bar{t}}\right)^{\frac{1}{4}}\,,
\end{equation}
where $\bar{t}=\eta_e/(\Delta\rho\omega^2V^{2/3})$ and the factor $\ln(t/\bar{t})$ arises from a leading-order expansion of $\ln(l/2a_0)$, which is assumed much larger than 1. It is worth noting that for our drops with 5\% wt H\textsubscript{2}O and 95\% wt glycerol in a bath of pure glycerol one has $\bar{t}=$ 0.032 s and the logarithmic correction is negligible on the typical experimental time scale.

For a larger viscosity ratio, $\lambda\gg(l/2a_0)^2/ \ln(l/2a_0) \gg 1$, extended drops resist deformation under the centrifugal pressure primarily owing to the internal gradient of the axial velocity. The deformation is analogous to the stretching of a piece of toffee \cite{lister_time-dependent_1996}. This case is not relevant to our experiments and will not be discussed further.


\section{Materials and methods}\label{matmeth}
\subsection{Materials}
Glycerol ($\geq$99.5\% wt) was purchased from Sigma Aldrich and used without further purification.
Silicon Oil (SO) has been purchased from Brookfiled Ametek and used as received. Milli-Q ultra pure water has been employed to prepare the water-glycerol mixtures. Densities $\rho$ and zero-shear viscosities $\eta$ at T=25 $^\circ$C of these liquids are reported in Table \ref{tab:table0}.
Fluorescein (disodium salt) (from Merck KGaA) was dissolved in all water/glycerol drops (at a concentration of 2$\cdot$ 10$^{-3}$ wt/wt), for which pure glycerol was the background fluid.\\
\begin{table}\tlstyle
	\begin{center}
		\centering
\begin{threeparttable}
		
		\begin{tabular}{c|c|c}
\toprule
				Liquid & $\rho$ (g/cm$^3$) & $\eta$ (mPa s)\tnotex{tnote:robots-r1}\\
				\midrule
				SO & 0.971\tnotex{tnote:robots-r2} & 91680\\
				Glycerol ($c_w\leq$0.02) & 1.260 $\pm$ 0.001 & 800.0 $\pm$ 0.1\\
				H$_2$O-Gly ($c_w$=0.05)& 1.250 $\pm$ 0.001\tnotex{tnote:robots-r3} & 430.8 $\pm$ 0.1
\\
\bottomrule		
\end{tabular}

\begin{tablenotes}
\item\label{tnote:robots-r1}The viscosity of the water-glycerol mixture and the pure glycerol were measured performing steady rate rheology experiments at shear rates ranging from 10 s$^{-1}$ up to 300 s$^{-1}$, using a stress-controlled AR 2000 rheometer (TA Instruments) equipped with a steel cone-and-plate geometry (cone diameter = 50 mm, cone
angle = 0.0198 rad). All samples showed pure Newtonian response with no dependence of viscosity on the shear rate.
\item\label{tnote:robots-r2} The viscosity and the density of SO are those quoted by the supplier company.
\item\label{tnote:robots-r3} The density of the water-glycerol mixture was inferred from the measured zero-shear viscosity, using tables reporting both $\eta$ and $\rho$ as a function of $c_w$ \cite{noauthor_physical_1963}.
\end{tablenotes}

\end{threeparttable}
\caption{\label{tab:table0} Mass density and viscosity of the fluids used in this work.}
	\end{center}
\end{table}

\subsection{Experimental setup}
All experiments were performed with a Kr\"{u}ss spinning drop tensiometer. Rates of rotation were accurate to 1\%.
The temperature was always set to 25.0$\pm$0.5$^\circ$C and kept constant using a flow of temperature-controlled air.
Different tests were performed with rotation rates ranging from 6000 rpm to 15000 rpm, such that buoyancy effects were negligible compared to centrifugal ones: the displacement of the drop off the rotation axis due to buoyancy was always smaller than 5 $\mu$m, as calculated following Ref. \cite{currie_buoyancy_1982}. All drops were illuminated by a blue Light Emitting Diode (LED) with a dominant emission wavelength of 469 nm. Measurements were performed using a cylindrical capillary with internal diameter $2R_c$=3.25 mm.
A CMOS camera (Phantom Miro 310 by Vision Research) run at 100 frames per second was used to record movies during the relaxation of immiscible drops of SO in glycerol. The camera was equipped with an objective by Nikkor (AF-Micro Lens 60 mm f/2.80, yielding a magnification $M = 1$).
No dye labelling was used for the immiscible fluids, since the persistent, sharp contrast of refractive index between the drop and the surrounding fluid allows the drop boundary to be clearly visualized (see Fig. \ref{Fig1} below) and its position to be accurately tracked in time, with an accuracy of 1 pixel, equivalent to 20 $\mu$m.
For the immiscible SO/glycerol systems, the lighter fluid (SO) was deposited on one cap of the capillary. The capillary was spun at high speed and $\omega$ was then reduced by an amount $\Delta\omega$. This resulted in the formation of an isolated drop due to the viscous pinch-off mechanism induced by surface tension \cite{stone_experimental_1986}. Drops of different volume $V$ were obtained by depositing a larger initial volume of SO in the capillary.

\begin{figure}
  \includegraphics[width=12cm]{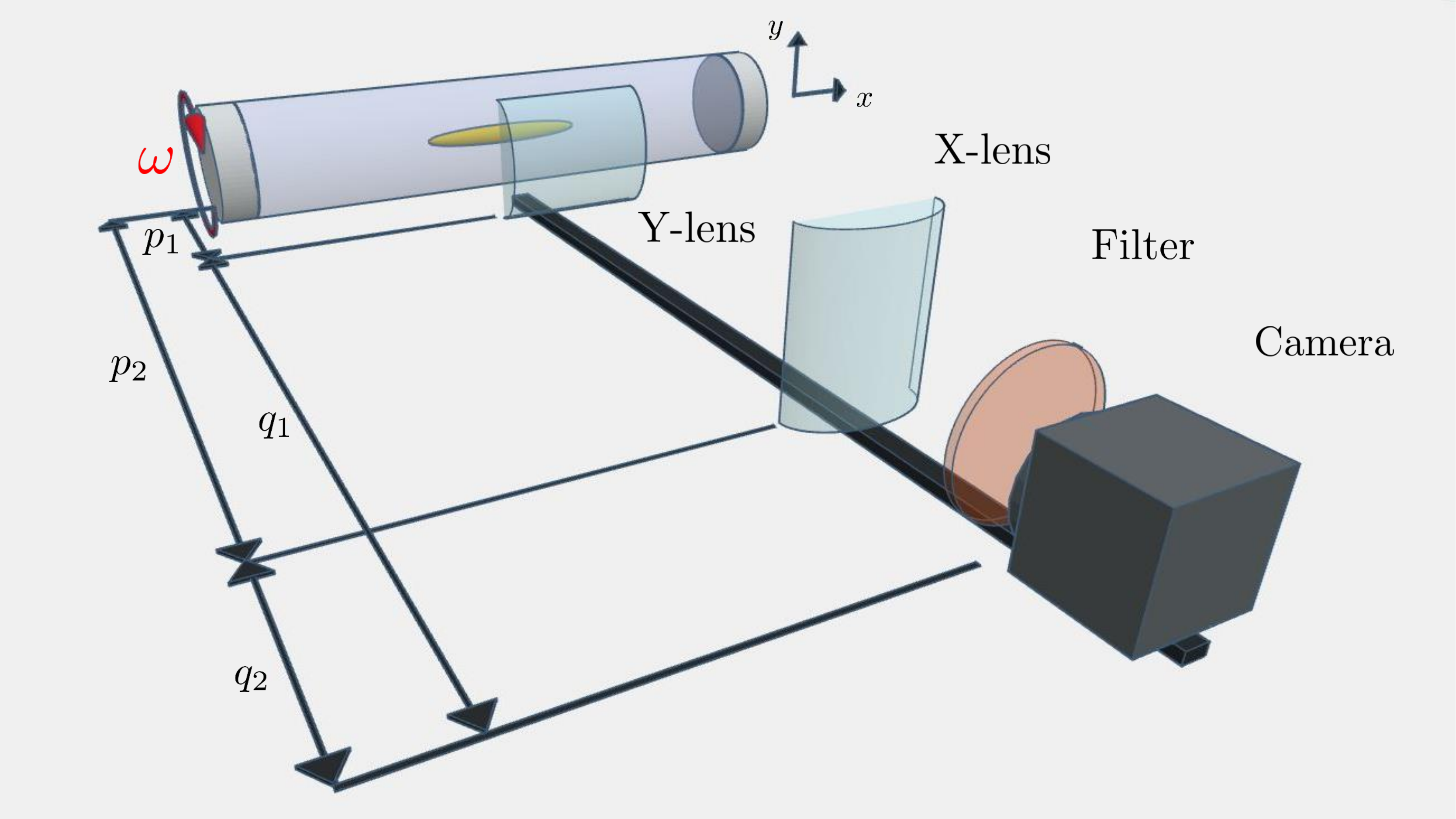}\\
  \caption{Setup used to characterize the extensional dynamics of fluorescent drops (water-glycerol mixture, $c_w=0.05$) in a miscible background fluid (pure glycerol, $c_w\approx0$). The drop is retro-illuminated by a series of LED (not shown for clarity).}
  \label{Sketchsetup}
\end{figure}

Imaging the ``interface'' between two miscible fluids, by contrast, is more challenging. First of all, the sharp optical contrast characterizing the boundary between the two fluids when they are initially brought in contact vanishes in few seconds, due to diffusion that smears out the concentration gradient. The smearing time is much smaller than the typical duration of our experiments, which is on the order of several minutes. Fluorescent labelling was therefore needed to track the drop elongation. Under the illumination of the blue LED light, fluorescein-rich drops appear as bright green-yellow regions, since the fluorescein adsorption and emission spectra (in polar solvents) are peaked at $\lambda\approx$ 485 nm and $\lambda\approx$ 511 nm \cite{panchompoo_one-step_2012,szalay_effect_1964}, respectively. The good contrast with the dark background allows for a precise measurement of the position and shape of the miscible drops as a function of time. In particular, we measure the drop length from the intensity profile on the drop axis: we define $l$ as the distance between the two apical points of the drop endcaps where the intensity profile reaches half of its maximum value.

In order to follow the evolution of very elongated miscible drops, we have extended the field of view in the direction of the capillary axis (the $x$ direction in Fig. \ref{Sketchsetup}). The custom imaging setup designed to this end is sketched in Fig. \ref{Sketchsetup}. The setup consists in a set of two plano-convex cylindrical lenses (Newport CKX17-C) sharing the same optical axis. The first lens, labelled Y-lens in Fig. \ref{Sketchsetup}, expands the image in the vertical direction with magnification $M_y$ = 3.36, while the second one (X-lens) contracts it the horizontal direction, with magnification $M_x$ = 0.3. Both lenses have an effective focal lens of 7.5 mm, and distances $p_1=q_2$ = 9.5 cm and $p_2=q_1$ = 32 cm as indicated in Fig. \ref{Sketchsetup}.
With this configuration the field of view in the $x$ direction is 3.25 cm (half of the capillary length) while it is of 3 mm in the $y$ direction. The setup allows the fluorescein concentration profile to be measured accurately in the $y$ direction, while retaining a large field of view in the $x$ direction, as required to follow very elongated drops. A blue-light filter removes the background light emitted by the source. In this configuration, a B/W CMOS camera (Toshiba Teli BU406M) run at a frame rate $\leqslant$ 90 Hz was used, the image size being 2048 x 2048 pixels. The magnifications $M_x$ and $M_y$ were measured by inserting a calibrated plastic grid into the capillary filled with the external fluid (glycerol). The lattice spacing was then measured in both the $x$ and $y$ directions, yielding calibration constants of 63 pixel/mm and 730 pixel/mm in the $x$ and $y$ directions, respectively.

For miscible fluids, all drops were injected directly in the capillary flushed and pre-filled with glycerol. Care was taken to make sure that air bubbles were removed from the capillary before sample injection. All drops were formed in the pre-filled capillary at rest, using a 1$\mu$l syringe, and their volume has been fixed to 1 $\mu$l. The microsyringe was smoothly removed from the capillary, to avoid the formation of tails of fluoresceinated water-glycerol mixtures that would have hampered the formation of initially quasi-spherical drops.

In the following we shall compare our experimental results to theories assuming that, upon a change of $\omega$, the final rotational speed is attained instantaneously and the fluids move solidly with the capillary. In experiments, however, changing $\omega$ and attaining a new rotational speed, uniform everywhere in the capillary, require a finite time due to the setup inertia and finite momentum diffusivity. It is therefore important to estimate these time scales and compare them to that of the drop evolution. In order to estimate $\tau_{\mathrm{setup}}$, the time required for the setup to change $\omega$, we consider the worst-case scenario, $\omega_i = 6000$ rpm and $\omega_f = 15000$, and measure the relaxation dynamics of a fast-relaxing drop of butanol in water, for which the theoretical relaxation time according to Eq.~\ref{Stonerelax} is $\tau\ll 0.1$ s. We measure a much larger relaxation time of about 0.3 s, which we thus identify with the setup response time $\tau_{\mathrm{setup}}$. The time for the system (capillary, drop, and background fluid) to move uniformly at one single $\omega$ can be estimated as the ratio between the squared capillary radius and the momentum diffusivity. Typical values are of the order of $\tau_{\mathrm{diff}} \simeq\frac{R_0^2\rho}{\eta_e}\approx 0.01$ s. In our experiments on both miscible and immiscible fluids, the drops always evolve on time scales much larger than both $\tau_{\mathrm{setup}}$ and $\tau_{\mathrm{diff}}$; thus, we shall consider that the final rotation speed and the rigid motion of the fluids are attained instantaneously.

\section{Drop extension in immiscible liquids}\label{immiscible}
In order to test the predictions of existing theories, we investigated the relaxation dynamics of SO drops in an immiscible glycerol background, upon a rotation speed jump.

\begin{figure}[htbp]
  \centering
  \includegraphics[width=6.5cm]{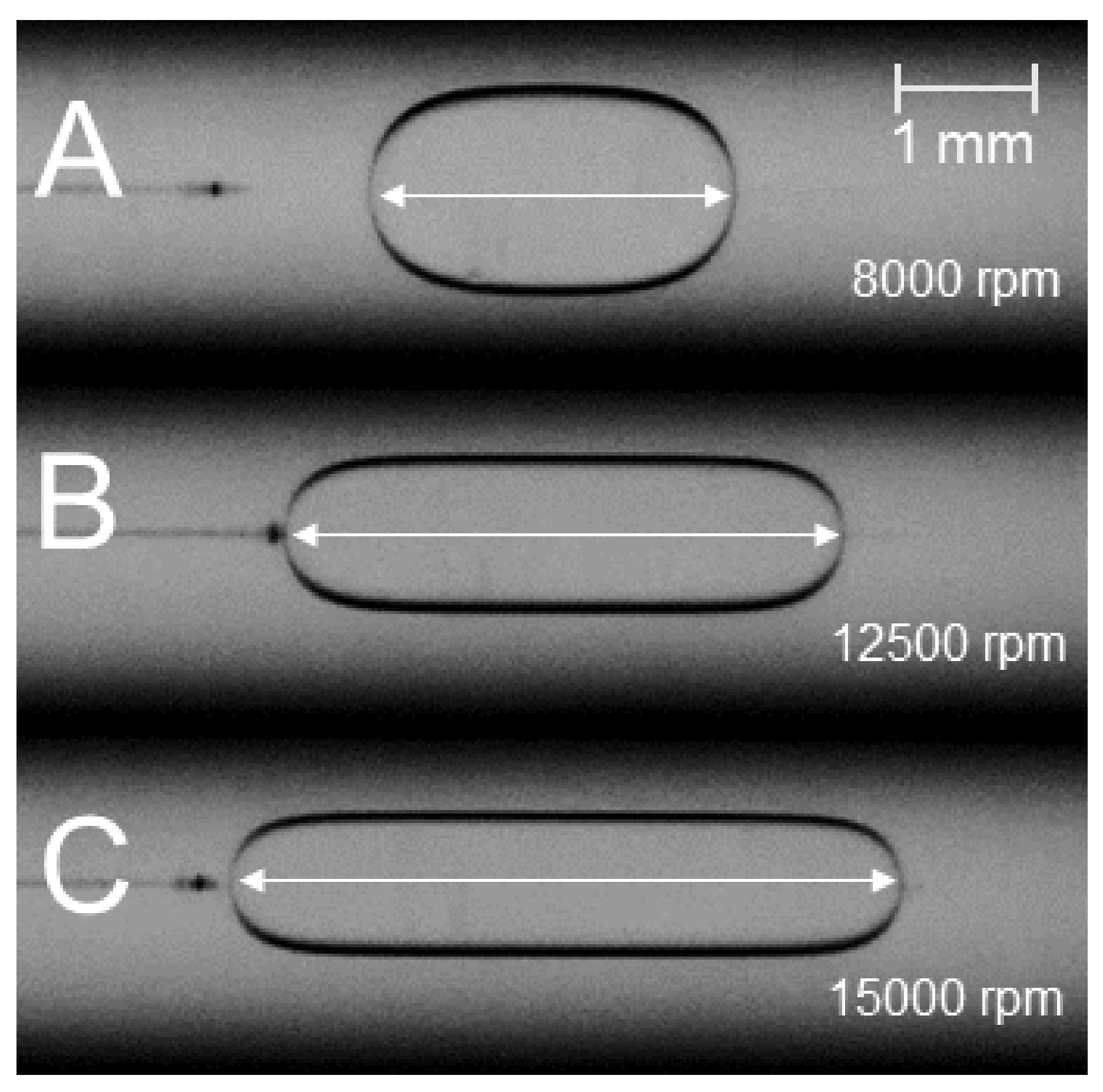}\\
  \includegraphics[width=8.5cm]{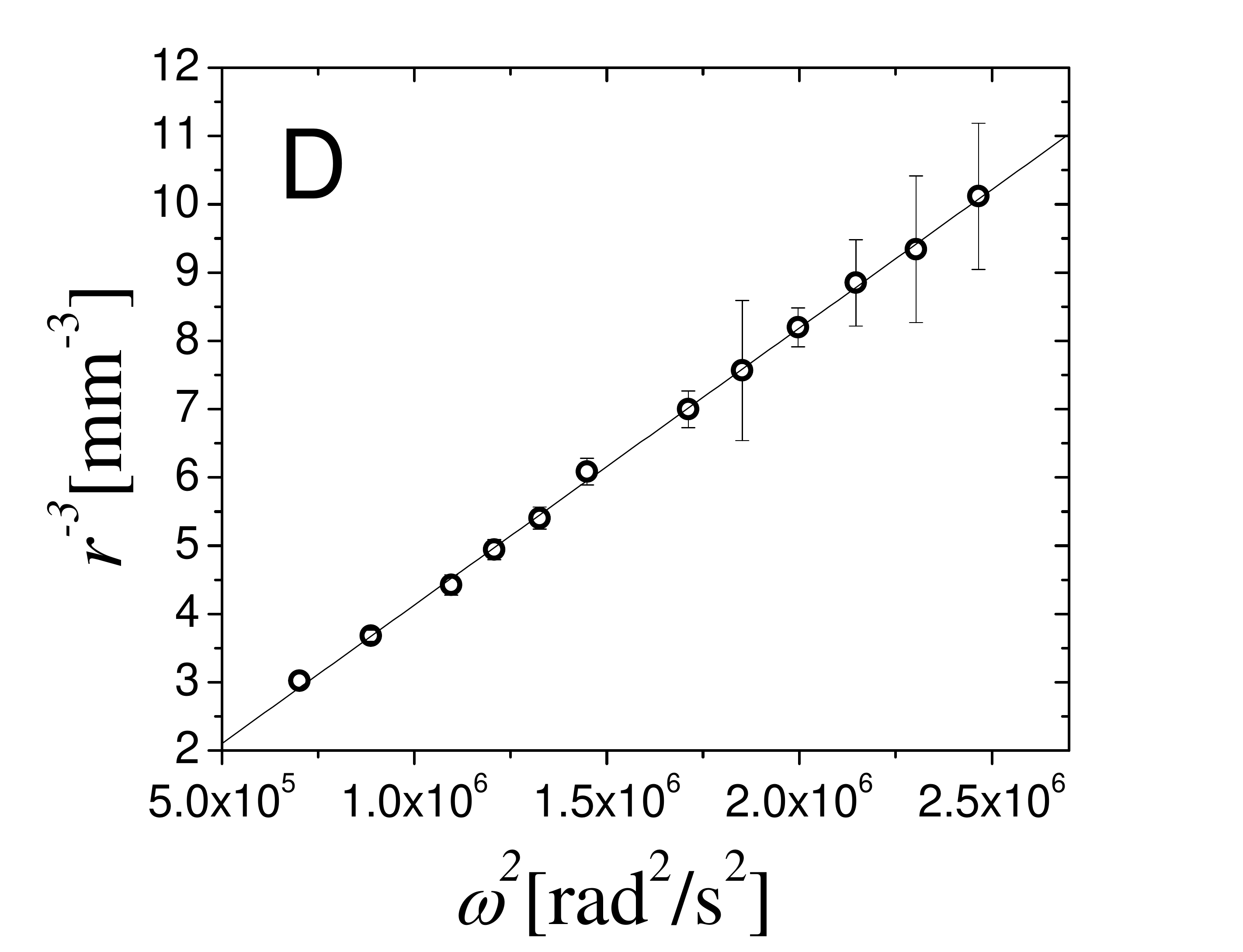}\\
  \caption{Spinning drops of SO at equilibrium in a reservoir of glycerol. A: $\omega=$ 8000 rpm. B: $\omega=$ 12500 rpm. C: $\omega=$ 15000 rpm. The white arrows show the length of the drops. D: $r^{-3}$ vs $\omega^2$ for one SO drop of volume $V_1$=2.95 $\mu$l in glycerol. The equilibrium radius is measured at the midsection of the drop. The solid line is a linear fit of the data, from which $\Gamma$ has been calculated using Vonnegut's formula (Eq. \ref{Vonnegut}). Error bars are calculated from the standard deviation of the radius over time.}\label{Fig1}
\end{figure}

We start by showing in panels A to C of Fig. \ref{Fig1} the equilibrium shape of a 2.95 $\mu$l SO drop in glycerol at different rotation speeds $\omega$. In agreement with Vonnegut's and Laplace's theories \cite{viades-trejo_spinning_2007}, we observe an increased stretching of the drop as $\omega$ grows. As predicted by Vonnegut's theory \cite{vonnegut_rotating_1942,viades-trejo_spinning_2007} for drops with aspect ratio $l/r \gtrsim 4$, we observe a direct proportionality between $r^{-3}$ and $\omega^2$ once the equilibrium shape of the drops is attained (Figure \ref{Fig1}-D). From the slope of the linear regression we obtain the surface tension between SO and glycerol: $\Gamma=$(17.81 $\pm$ 0.02) mN/m. It's worth noting that this value is less than half of that reported for SO and water ($39.8$ mN/m \cite{koos_tuning_2012}) and is in agreement with recent results obtained for $\Gamma$ in systems of nonpolar mineral oils in contact with water-glycerol mixtures with variable glycerol fractions \cite{sinzato_experimental_2015}. We emphasize that by fitting $r^{-3}$ vs $\omega^2$ one reduces the experimental uncertainty affecting $\Gamma$ with respect to measuring the drop radius at one single rotation speed. While yielding a very precise measurement, this procedure reduces the complications inherent to methods relying on the detection of the whole drop shape, e.g. the Young-Laplace fit of the drop surface, which are more sensitive to refraction effects due to the cylindrical capillary and are affected by any change in the illumination conditions.

\begin{figure}[htbp]
  \centering
  \includegraphics[width=10cm]{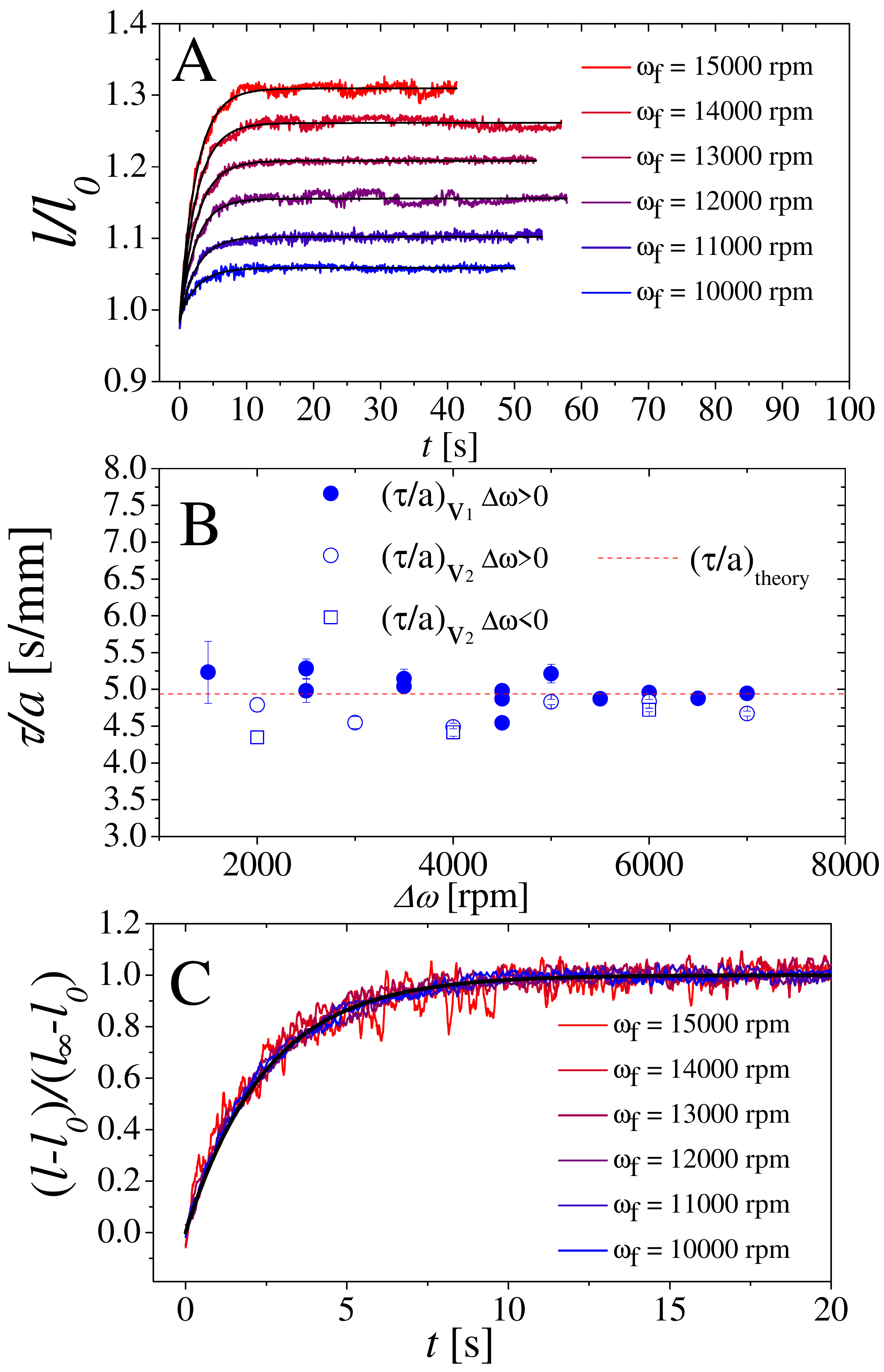}\\
  \caption{A: Normalized length $l/l_0$ of silicon oil drops in glycerol as a function of time for different rotation speed jumps ($\omega_f-\omega_i$), with $\omega_i$ = 8000 rpm. B: Relaxation time normalized by the drop radius at rest as a function of the magnitude of the speed jump, for all experiments. The drop volume and the sign of $\Delta\omega$ are shown by the labels. The dashed red line shows the prediction of Eq. \ref{Stonerelax} \cite{stone_stone_1996}. C: Scaled length $(l-l_0)/(l_\infty-l_0)$, where the initial, $l_0$, and asymptotic, $l_\infty$, drop lengths are obtained as detailed in the text.}\label{Fig2}
\end{figure}

The relaxation of the drop towards its equilibrium shape following a sudden jump of rotation speed $\Delta \omega=\omega_f-\omega_i$ has been investigated in a series of tests. In most experiments, the drop volume is fixed to $V_1$= 2.95 $\mu$l and the initial speed is $\omega_i=$ 8000 rpm. The relaxation dynamics are measured for a series of rotational speed jumps ranging from $\Delta\omega$ = 1000 rpm to $\Delta\omega$ = 7000 rpm. The time dependent drop length normalized by $l_0$ is shown in Fig. \ref{Fig2}-A. Some of the relaxation tests have been repeated for smaller drops with volume $V_2$= 0.66 $\mu$l. Finally, we have performed drop retraction experiments, corresponding to $\Delta \omega<0$.

Equation \ref{Stonerelax} predicts that the relaxation time $\tau$ normalized by the drop radius at rest $a$ is independent of the drop volume $V$ and of $\Delta\omega$, while $\tau$ should depend on the value of the interfacial tension and on the viscosities of the two fluids. We test this prediction by showing all the normalized relaxation times $\tau/a$ in Fig. \ref{Fig2}-B. Indeed, the measured normalized relaxation times are neither affected by the drop volume, nor by the magnitude or the sign of $\Delta \omega$. To further corroborate this scenario, we report in Fig. \ref{Fig2}-C the normalized relaxation curves $\frac{l-l_0}{l_{\infty}-l_0}$, where $l_{\infty}$ and $l_0$ are measured at equilibrium for each rotation speed. All data collapse onto the mastercurve $\frac{l-l_0}{l_{\infty}-l_0}=1-\exp[-(t/\tau)_{av}]$, with $(\tau/a)_{av}=4.9 \pm 0.2$ s/mm the scaled relaxation time averaged over all the experiments.



We find an excellent agreement between the scaled drop relaxation time and the theoretical value predicted by the Stone and Bush theory \cite{stone_stone_1996}, $(\tau/a)_{th}=\frac{\eta_e}{\Gamma}f(\lambda)=$ 4.97 $\pm$ 0.01 s/mm. Remarkably, the theory of Ref. \cite{stone_stone_1996} turns out to be an excellent predictive tool for the relaxation dynamics of viscous drops well beyond the limit of nearly spherical drops and small Bond number $Bo=\frac{\Delta\rho\omega^2 a^3}{\Gamma}\ll$1, for which it has been derived. Indeed, for our SO drops the aspect ratio $l/2r$ ranges from 1.00 to 5.25 and 1.4$\leq Bo\leq$9.1. Our findings suggest that higher order spherical harmonics that should be added to model the shape of elongated droplets, leading to corrections to the expression derived in Ref. \cite{stone_stone_1996} for the velocity and pressure fields, do not influence appreciably the drop dynamics. More theoretical work will be needed to confirm this scenario.


\section{Evolution of drops in a miscible background fluid}\label{miscible}
\begin{figure}[htbp]
  \includegraphics[width=7.5cm, height=6.5cm]{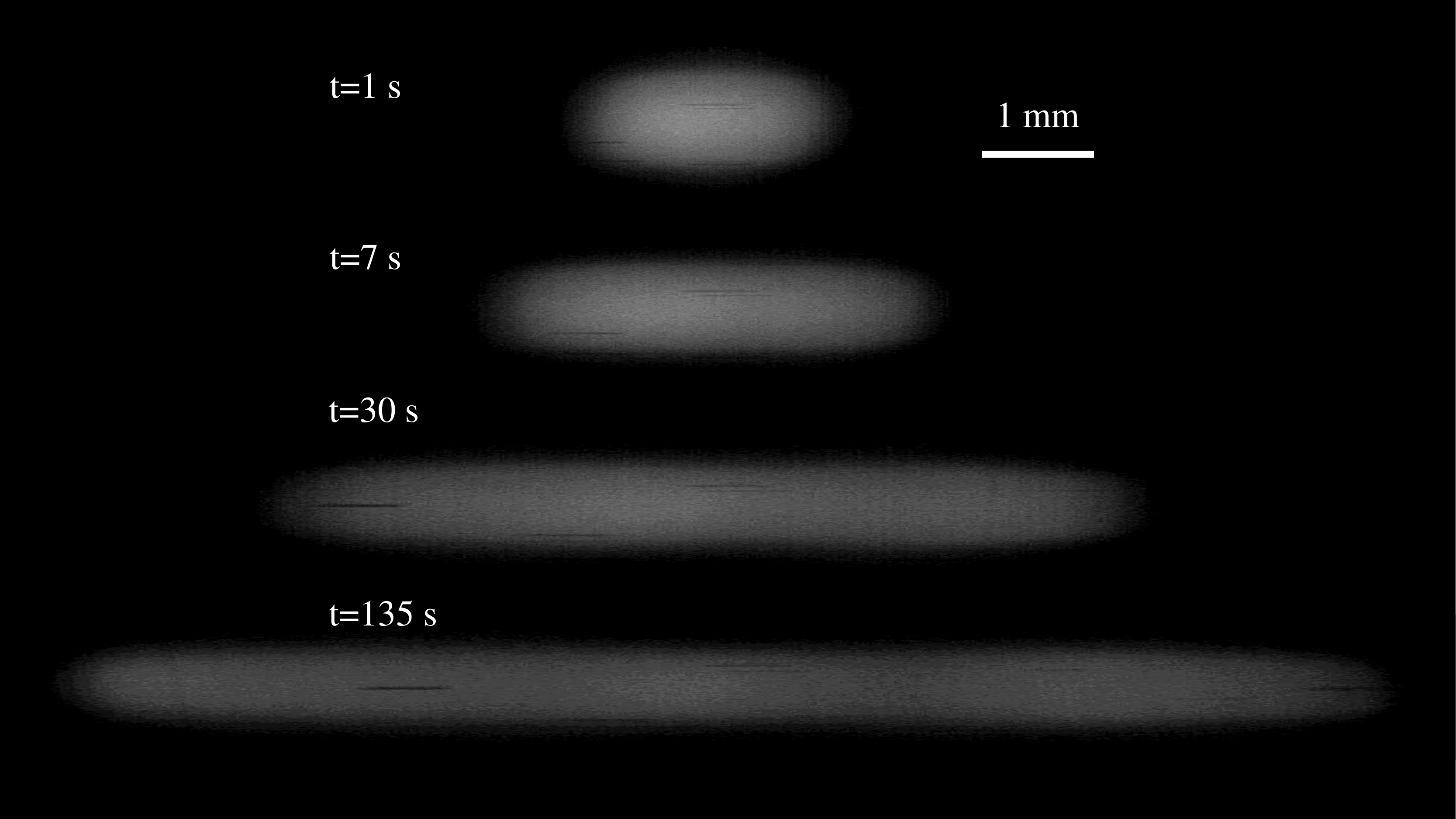}\\
  \caption{Temporal evolution of the H$_2$O-Glycerol drop ($c_w$=0.05) in pure glycerol at $\omega_f$=15000 rpm. $t$=0 represents the beginning of the rotation. All drops are shown at an aspect ratio of 1.}\label{Evolmisc}
\end{figure}

Studying the drop evolution in a miscible background fluid is not an easy task as capillary effects due to a gradient of concentrations \cite{truzzolillo_off-equilibrium_2017,truzzolillo_nonequilibrium_2016} are transient and their magnitude is expected to be very weak in miscible molecular fluids, for which typical estimates of the EIT are smaller than 1 mN/m \cite{petitjeans_petitjean_1996,pojman_evidence_2006,zoltowski_evidence_2007}. Moreover, stresses induced by secondary flows in the spinning capillary may be comparable to or even larger than those due to surface tension, resulting in permanent drop deformation or time dependent instabilities \cite{manning_interfacial_1977}. Indeed, when a capillary starts rotating, the background fluid within the shear layer is pushed towards the drop center, as pointed out by Currie et al. \cite{currie_buoyancy_1982} and Manning et al. \cite{manning_interfacial_1977}, who showed that this effect is large enough to deform drops in water-hydrocarbon-surfactant systems with an equilibrium interfacial tensions of the order of 10 $\mu$N/m. Secondary flows cause long lasting dog-bone drop shapes that are unusable for determining the interfacial tension with Eq. \ref{Vonnegut}, and affect the drop elongation dynamics. For our miscible systems, we have observed dog-bone shapes for all drops composed of a water-glycerol mixture with a water mass fraction $c_w>0.05$ and immersed in pure glycerol. We shall discuss the evolution of these dog-bone shaped drops in a forthcoming publication. Here, we report data for drops with $c_w=0.05$, which retain a end-capped cylindrical shape over (at least) thousands of seconds, for all probed $\omega_f$. A typical drop is shown in Fig.~\ref{Evolmisc}, for different times $t$ following a jump of the rotational speed from $\omega_i = 0$ to $\omega_f = 15000~\mathrm{rpm}$ at $t=0$.

\begin{figure}[htbp]
  \includegraphics[width=10cm]{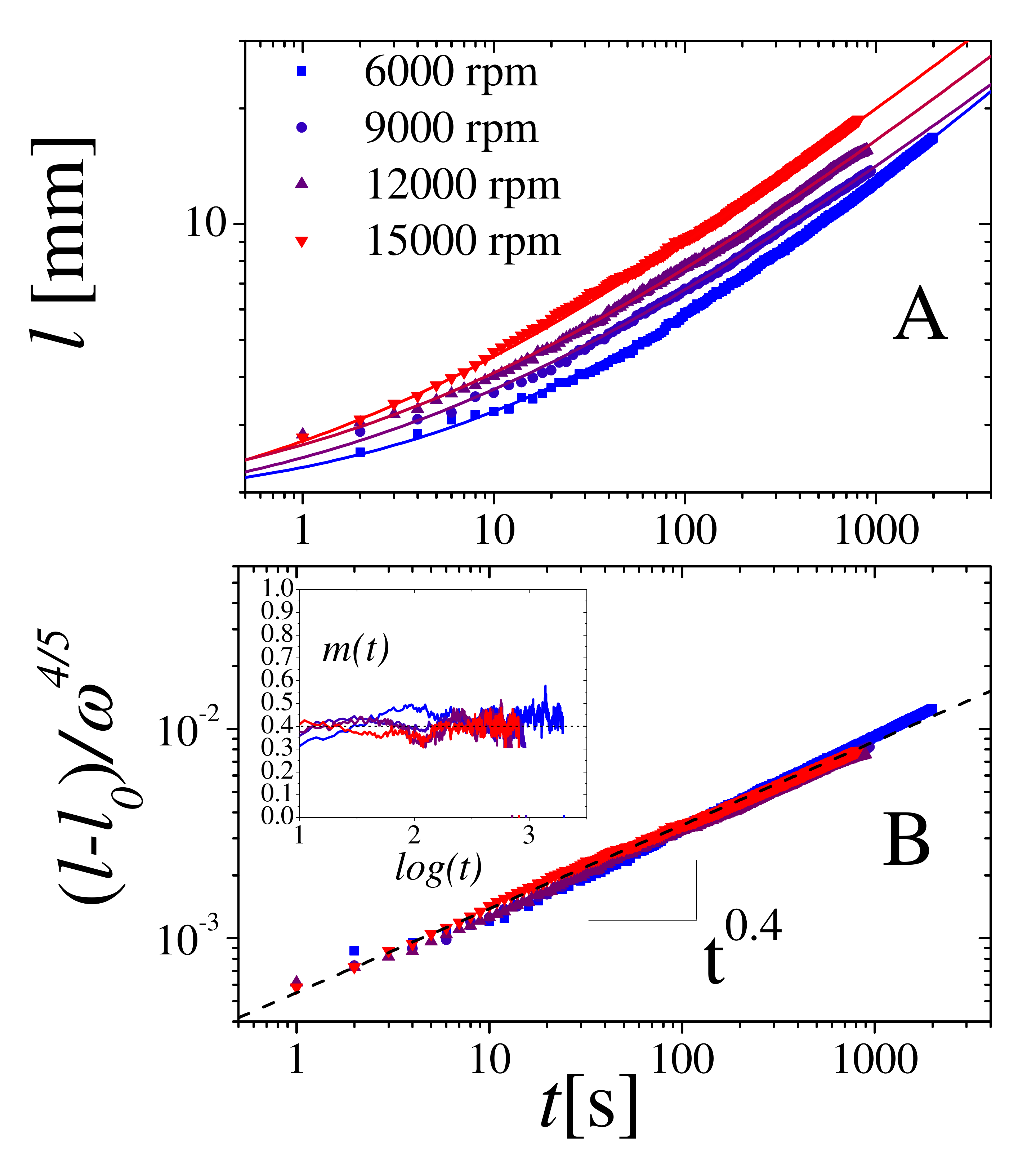}\\
  \caption{A: Temporal evolution of the H$_2$O-Glycerol drop length in pure glycerol at different rotation speed $\omega_f$ as indicated by the labels. Solid lines are best fits to a power-law growth, $l(t)=l_0+\alpha t^m$.
  B: Rescaled increase of the drop length $(l-l_0)/\omega_f^{4/5}$, as a function of time. The dashed line shows the best fit obtained using the scaling law $l-l_0\sim t^{\frac{2}{5}}$ predicted by Lister and Stone \cite{lister_time-dependent_1996} (Eq. \ref{lenghtmisc1}). The inset shows the time-dependent (local) exponent $m(t)=\mathrm{d}\log(l-l_0)/\mathrm{d}\log(t)$.}\label{Scalingmisc}
\end{figure}

Figure \ref{Scalingmisc}-A shows the drop length as a function of time for four identical drops, with $V=1~\mu\mathrm{ l}$ and $c_w=0.05$. In all experiments, $\omega_i = 0$, while the final angular speed has been varied in the range $6000$ rpm - $15000$ rpm, as shown by the labels. In all cases, the increase of the drop length is very well fitted by a power-law: $l(t)=l_0+\alpha t^m$. The fitting parameters $\alpha$ and $m$ are reported in Table~\ref{tab:table1}. To monitor possible deviations from a power-law behaviour, we compute the time-dependent (local) exponent $m(t)=\frac{\mathrm{d}log[l(t)-l_0]}{\mathrm{d}log(t)}$, shown in the inset of Fig.~\ref{Scalingmisc}-B. For all datasets, $m$ is very close to 0.4 throughout the duration of the experiment, with no systematic deviations.

\begin{table}
	\begin{center}
		\centering\tlstyle
  \begin{threeparttable}
		\begin{tabular}{c|c|c}
\toprule
				$\omega_f$ (rpm) & $\alpha$ (mm/s$^m$) & $m$ \tnotex{tnote:robots-r1}\\
				\midrule
				15000 & 1.227 $\pm$ 0.003 & 0.3934 $\pm$ 0.0004\\
				12000 &  0.962 $\pm$ 0.003 & 0.3960 $\pm$ 0.0005\\
				9000 & 0.864 $\pm$ 0.002 & 0.3870 $\pm$ 0.0005\\
				6000 & 0.525 $\pm$ 0.001 & 0.44 $\pm$ 0.02\\
\bottomrule
		\end{tabular}
\begin{tablenotes}
      \item\label{tnote:robots-r1}The reported errors are those on the fit parameters of the non-linear regression of the experimental data performed with the Origin software (OriginLab corporation)
    \end{tablenotes}
\end{threeparttable}
\caption{\label{tab:table1} Fitting parameters for the data shown in Fig.~\ref{Scalingmisc}.}
	\end{center}
\end{table}
The values obtained for $m$ are in very good agreement with $m=\frac{2}{5}$ as predicted by Lister and Stone for the ``bubble-like'' regime (Eq.~\ref{lenghtmisc1}), e.g. for drops with negligible effects of intra-drop stresses. To further test the applicability of Eq.~\ref{lenghtmisc1}, we inspect the $\omega_f$ dependence of the the prefactor $\alpha$. We find $\alpha \sim \omega_f^{0.83 \pm 0.12}$, very close to the scaling $\alpha\sim\omega_f^\frac{4}{5}$ predicted by Eq.~\ref{lenghtmisc1}. Indeed, Fig.~\ref{Scalingmisc}-B shows that all the data nicely collapse onto a mastercurve when using the scaled variable $(l-l_0)/\omega_f^{4/5}$ as suggested by Eq.~\ref{lenghtmisc1}. Interestingly, we find that the bubble-like regime extends beyond the limits described originally in Ref.~\cite{lister_time-dependent_1996}. Indeed, in our experiments $\lambda=0.537 < 1$ and the aspect ratio of the drop at its maximum (measurable) extension is $2a_0/l \approx 0.02$, so that one has $2a_0/l \ll \lambda < 1$. This is to be compared to $\lambda \ll 2a_0/l \ll 1$, as postulated in Ref.~\cite{lister_time-dependent_1996} for the bubble-like regime.

Overall, our data for miscible drops are very well accounted for by the model of Ref.~\cite{lister_time-dependent_1996}, which was developed for immiscible fluids in the limit $\Gamma \rightarrow 0$. The agreement with the theory by Lister and Stone strongly suggests that capillary forces do not affect significantly the dynamics of our miscible drops. This is consistent with what expected for a pair of miscible fluids with a very low compositional gradient at their boundary (due to molecular diffusion). It is furthermore consistent with previously reported data for the effective interfacial tension in water-glycerol mixtures~\cite{petitjeans_petitjean_1996}, from which we expect $\Gamma \ll 0.02~\mathrm{mN/m}$ for our system.

It is worth noting that the interface between the two fluids is supposed to be infinitely sharp in Ref.~\cite{lister_time-dependent_1996}, while it is smeared out by diffusion in our experiments. It is thus important to analyze the effects of diffusion, in order to rationalize why the diffusion-free model by Lister and Stone describes so well our data. Due to diffusion, the effective surface tension tends to decrease with time, since the composition gradient at the interface diminishes. However, this evolution is most likely masked by the fact that for miscible systems $\Gamma$ is anyway low enough to meet the $\Gamma \rightarrow 0$ condition postulated in Ref.~\cite{lister_time-dependent_1996} at all times. Diffusion may also impact the drop evolution in a more subtle way, by coupling to the centrifugal forcing: the forcing may change the evolution of the concentration profile; in turn, a smeared profile modifies the centrifuge forces. We first show that in fact the centrifugal forcing does not affect Fickian diffusion within the capillary. The length scale over which the mass distribution of the two liquids is affected by rotation may be estimated by calculating the approximate equilibrium profile of water molecules in a water/glycerol mixture at $\omega > 0$. Following \cite{obidi_analytical_2017} we obtain \cite{noauthor_supplemental_nodate}:
\begin{equation}\label{exp-rotational}
  x_w(r)\sim \exp\left[-\frac{r^2}{l_{\omega_f}^2}\right] \,,
\end{equation}
where $x_w$ is the mole fraction of water in the mixture and $l_{\omega_f}$ is the rotational length
\begin{equation}\label{rotational-length}
  l_{\omega_f}=\frac{1}{\omega_f}\sqrt{\frac{2RT}{v_w\rho_{av}-M_w}} \,,
\end{equation}
i.e. the analogous to the gravitational length calculated under static conditions~\cite{obidi_analytical_2017}. Here $R$ is the universal gas constant, $T$ the absolute temperature, $M_w$ and $v_w$ the molecular weight and the partial molar volume of water and $\rho_{av}$ the average density of the mixture. For water and glycerol at $\omega_f=15000$ rpm we obtain $l_{\omega_f}\simeq$ 2.75 m much greater than the capillary radius \cite{noauthor_supplemental_nodate}, showing that in our system the centrifugal forcing is far from affecting inter-diffusion between the drop and the surrounding liquid.

Having established that diffusion can be safely taken to be Fickian, we discuss its effect on the centrifugal forcing. Although the diffusion of water initially contained in the drop is a slow process, its effect over the typical duration of our experiments (several hundreds of seconds) is not negligible. Indeed, it takes about 710 s for a water molecule to diffuse in glycerol over a distance of $100 ~\mu\mathrm{m}$~\cite{derrico_diffusion_2004}: diffusion does indeed smear out significantly the interface between the drop and the surrounding medium during the experiment. Because of smearing, the pressure field in the capillary is different from that for a sharp interface. In our experiments the dominant contribution to the drop dynamics arises from the pressure jump $\Delta P$ across the endcaps of the drop; we thus calculate $\Delta P$ for the two cases of a sharp or diffused interface. For a sharp interface, the net pressure exerted on the head of a quasi cylindrical drop is~\cite{lister_time-dependent_1996}
\begin{equation}\label{pjump1}
    \Delta P_{sharp}\simeq\frac{1}{2}\Delta\rho\omega_f^2r_0^2 \,,
\end{equation}
where $r_0$ is the radius of the drop with sharp boundaries.

\begin{figure}[htbp]
	\includegraphics[width=8cm]{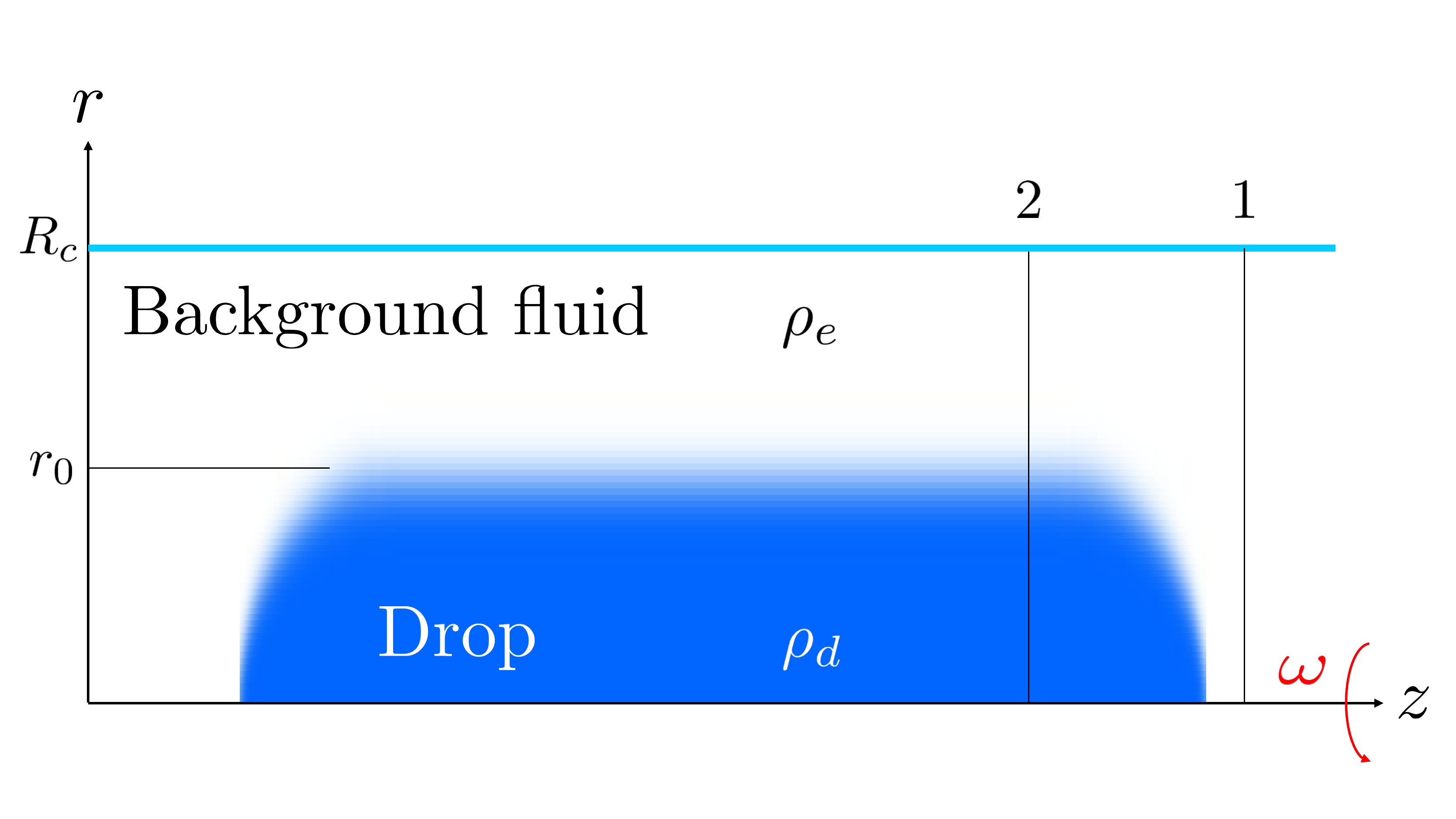}\\
	\caption{Scheme for the general case of a drop with diffused interface. The pressure field can be integrated on lines 1 and 2 to obtain the forcing on the drop.}\label{sketch_DeltaP}
\end{figure}

Equation~\ref{pjump1} is actually a particular case of a more general expression that arises from the difference between the pressure field integrated along two distinct radial trajectories: one lying outside the spinning drop, the second one partially included in the drop, as sketched in Fig.~\ref{sketch_DeltaP}. Under rotation, this gives rise to an effective hydrostatic pressure on the drop that reads:
\begin{equation}\label{pjump2a}
   \Delta P_{d} \simeq \omega_f^2\int_0^{R_c}\left[\rho_1(r)-\rho_2(r)\right]r \mathrm{d}r  \,,
\end{equation}
where $\rho_1$ is the density profile along the line 1, equal to the density of the surrounding liquid, $\rho_2$ is the density profile along the line 2 and $R_c = 1.62\mathrm{5~} \mathrm{mm}$ for our capillary. The density profile along line 2 is computed by solving the diffusion equation along the radial direction $r$.

Since diffusion is slow, one can simplify the problem by considering the section sketched in Fig.~\ref{sketch_DeltaP} as being composed by two semi-planes separated by an initially planar boundary at distance $r_0$ from the origin, the latter corresponding to the rotation axis of the capillary. Considering that the density of water-glycerol mixtures varies linearly with the water weight fraction for $0\leq c_w\leq0.05$~\cite{cheng_formula_2008}, Eq.~\ref{pjump2a} can be written as
\begin{equation}\label{pjump2b}
   \Delta P_{d} \simeq \omega_f^2\int_0^{R_c}\frac{\Delta \rho}{2}-\frac{\Delta \rho}{2}\erf\left[\frac{(r-r_0)}{\sqrt{4D_wt}}\right]r \mathrm{d}r \,,
\end{equation}
where $r_0$ is the drop radius, $\Delta\rho$=10 kg/m$^3$ is the initial density difference between the drop and the background fluid, $D_w=$1.4 $\cdot$ 10$^{-11}$ m$^2$/s is the diffusion coefficient of water in glycerol \cite{derrico_diffusion_2004} and $t$ is the time elapsed since the drop has been inserted in the capillary. From the experimental images we estimate that the radius $r_0$, defined as the position of the flex point of the radial intensity profile, lays between 0.25 and 0.35 mm. By using Eq.~\ref{pjump2b} an upper bound for the relative variation of the hydrostatic pressure can be calculated: $(\Delta P_{sharp}-\Delta P_{d})/\Delta P_{sharp}\leq 30\%$ for $t=1000$ s, while the same quantity is as small as 3\% for $t=100$ s. We thus expect that the dynamics are essentially not affected by diffusion-driven effects for the first few hundred seconds, while some limited impact may be present in the last time decade ($10^2$-$10^3$ s) explored in our experiments. Consistently with this analysis, Fig.~\ref{Scalingmisc} shows no significant deviations from the Lister-Stone theory up to $t=2000$ s. Note that, as pointed out first by Zeldovich~\cite{Zeldovich1949} and then by Zoltowski et al.~\cite{zoltowski_evidence_2007}, mass conservation implies that as the drop is stretched the transition zone at its boundary is thinned. This effect likely keeps the concentration profile sharper than expected from Fickian diffusion, thus contributing to maintain the scaling predicted by Eq.~\ref{lenghtmisc1}.

\section{Conclusions}\label{conclusions}
We have investigated experimentally the relaxation of Newtonian spinning drops after a jump of angular speed, detailing two radically different cases: i) immiscible fluids with viscosities high enough for the slow extensional dynamics to be fully uncoupled from the inertial response of the tensiometer; ii) miscible fluids with negligible capillary effects and no deformations due to secondary flows in the rotating capillary. A crucial experimental improvement was the use of a custom imaging geometry, which allowed us to follow the elongation of drops up to very large aspect ratios, a regime never probed experimentally before. By imaging the full drop, we were furthermore able to rule out the presence of shape instabilities for all the samples discussed here, and unambiguously ascertain whether or not a steady state was eventually reached.

For immiscible fluids, we have found that the drop dynamics is described remarkably well by predictions for a quasi-spherical drop subject to small deformations, for which the typical relaxation time does not depend on the forcing, i.e. the product $\omega^2\Delta\rho$, nor on the equilibrium length or shape of the drops. For these immiscible drops, the relaxation dynamics depend only on the interfacial tension $\Gamma$, the drop radius at rest, and the viscosity of the drop and background fluids. An excellent quantitative agreement is found with the theory proposed in Ref. \cite{stone_stone_1996}, with a larger than expected domain of applicability.

Drops in miscible environments exhibit a totally different relaxation dynamics, their length increasing indefinitely according to a power law, in agreement with predictions for ``bubble-like'' dynamics in absence of capillary effects: $l(t)-l(0)\sim \alpha\cdot t^{\frac{2}{5}}$, with $\alpha\sim\omega_f^{4/5}$ \cite{lister_time-dependent_1996}. A global theory that fully captures the experimental droplet relaxation under the assumption of a pure unsteady extension superposed to a rigid rotation is still missing, as also pointed out by Joseph et al. \cite{joseph_spinning_1992}. Our results suggest that existing models based on the calculation of the contribution of the dominant viscous stresses inside and outside the drop reproduce very well the experimental data. This observation should be useful for guiding future theoretical efforts.

Finally, it is worth to briefly compare our results in the light of previous works, notably Ref.~\cite{petitjeans_petitjean_1996}, which reported SDT measurements of the effective interfacial tension between a drop of water and water-glycerol mixtures as the background fluid. In these experiments, which are similar to those reported here for miscible fluids, the concentration $c_g$ of glycerol in the background fluid was systematically varied. It was observed that $\Gamma$ vanishes as $c_g$ decreases, becoming unmeasurable for $c_g<$ 40 \% wt. This is consistent with our result that the elongation dynamics of drops with composition very close to that of the background fluid is well described by Lister's and Stone's theory, where interfacial tension is neglected~\cite{lister_time-dependent_1996}. Indeed, in our system the composition mismatch is smaller than the smallest one for which Petitjean was still able to measure $\Gamma$ (compare $c_w = 5\%$ wt to $c_g =  40\%$ wt).

The results presented here should help clarifying the debate on the behavior of spinning drops in miscible and immiscible background fluids. We hope that they will also stimulate further theoretical work in this field.

\begin{acknowledgement}
We thank L. Isa for fruitful discussions and gratefully acknowledge support from the Centre national d'\'{e}tudes spatiales (CNES).
\end{acknowledgement}

\begin{suppinfo}
Calculation of the rotational length for water-glycerol mixtures.
\end{suppinfo}



\begin{mcitethebibliography}{47}
\providecommand*\natexlab[1]{#1}
\providecommand*\mciteSetBstSublistMode[1]{}
\providecommand*\mciteSetBstMaxWidthForm[2]{}
\providecommand*\mciteBstWouldAddEndPuncttrue
  {\def\EndOfBibitem{\unskip.}}
\providecommand*\mciteBstWouldAddEndPunctfalse
  {\let\EndOfBibitem\relax}
\providecommand*\mciteSetBstMidEndSepPunct[3]{}
\providecommand*\mciteSetBstSublistLabelBeginEnd[3]{}
\providecommand*\EndOfBibitem{}
\mciteSetBstSublistMode{f}
\mciteSetBstMaxWidthForm{subitem}{(\alph{mcitesubitemcount})}
\mciteSetBstSublistLabelBeginEnd
  {\mcitemaxwidthsubitemform\space}
  {\relax}
  {\relax}

\bibitem[Rayleigh(1914)]{rayleigh_equilibrium_1914}
Lord Rayleigh, The equilibrium of revolving liquid under capillary force.
  \emph{Phil. Mag.} \textbf{1914}, \emph{28}, 161\relax
\mciteBstWouldAddEndPuncttrue
\mciteSetBstMidEndSepPunct{\mcitedefaultmidpunct}
{\mcitedefaultendpunct}{\mcitedefaultseppunct}\relax
\EndOfBibitem
\bibitem[Janiaud \latin{et~al.}(2000)Janiaud, Elias, Bacri, Cabuil, and
  Perzynski]{janiaud_spinning_2000}
Janiaud,~E.; Elias,~F.; Bacri,~J.-C.; Cabuil,~V.; Perzynski,~R. Spinning
  ferrofluid microscopic droplets. \emph{Magnetohydrodynamics} \textbf{2000},
  \emph{36}, 301--304\relax
\mciteBstWouldAddEndPuncttrue
\mciteSetBstMidEndSepPunct{\mcitedefaultmidpunct}
{\mcitedefaultendpunct}{\mcitedefaultseppunct}\relax
\EndOfBibitem
\bibitem[Hill and Eaves(2008)Hill, and Eaves]{hill_nonaxisymmetric_2008}
Hill,~R. J.~A.; Eaves,~L. Nonaxisymmetric {Shapes} of a {Magnetically}
  {Levitated} and {Spinning} {Water} {Droplet}. \emph{Physical Review Letters}
  \textbf{2008}, \emph{101}\relax
\mciteBstWouldAddEndPuncttrue
\mciteSetBstMidEndSepPunct{\mcitedefaultmidpunct}
{\mcitedefaultendpunct}{\mcitedefaultseppunct}\relax
\EndOfBibitem
\bibitem[Chandrasekhar(1961)]{chandrasekhar_hydrodynamic_1961}
Chandrasekhar,~S. \emph{Hydrodynamic and {Hydromagnetic} {Stability}}; Oxford
  University Press, 1961\relax
\mciteBstWouldAddEndPuncttrue
\mciteSetBstMidEndSepPunct{\mcitedefaultmidpunct}
{\mcitedefaultendpunct}{\mcitedefaultseppunct}\relax
\EndOfBibitem
\bibitem[Li \latin{et~al.}(2019)Li, Fang, Li, Yang, Li, Li, Feng, and
  Song]{li_spontaneous_2019}
Li,~H.; Fang,~W.; Li,~Y.; Yang,~Q.; Li,~M.; Li,~Q.; Feng,~X.-Q.; Song,~Y.
  Spontaneous droplets gyrating via asymmetric self-splitting on heterogeneous
  surfaces. \emph{Nature Communications} \textbf{2019}, \emph{10}\relax
\mciteBstWouldAddEndPuncttrue
\mciteSetBstMidEndSepPunct{\mcitedefaultmidpunct}
{\mcitedefaultendpunct}{\mcitedefaultseppunct}\relax
\EndOfBibitem
\bibitem[Filali \latin{et~al.}(2018)Filali, Er-Riani, and
  El Jarroudi]{filali_deformation_2018}
Filali,~Y.; Er-Riani,~M.; El Jarroudi,~M. The deformation of a ferrofluid drop
  under a uniform magnetic field. \emph{International Journal of Non-Linear
  Mechanics} \textbf{2018}, \emph{99}, 173--181\relax
\mciteBstWouldAddEndPuncttrue
\mciteSetBstMidEndSepPunct{\mcitedefaultmidpunct}
{\mcitedefaultendpunct}{\mcitedefaultseppunct}\relax
\EndOfBibitem
\bibitem[K\'{e}kesi \latin{et~al.}(2016)K\'{e}kesi, Amberg, and
  Prahl~Wittberg]{kekesi_drop_2016}
K\'{e}kesi,~T.; Amberg,~G.; Prahl~Wittberg,~L. Drop deformation and breakup in
  flows with shear. \emph{Chemical Engineering Science} \textbf{2016},
  \emph{140}, 319--329\relax
\mciteBstWouldAddEndPuncttrue
\mciteSetBstMidEndSepPunct{\mcitedefaultmidpunct}
{\mcitedefaultendpunct}{\mcitedefaultseppunct}\relax
\EndOfBibitem
\bibitem[Holgate and Coppins(2018)Holgate, and Coppins]{holgate_shapes_2018}
Holgate,~J.~T.; Coppins,~M. Shapes, stability, and hysteresis of rotating and
  charged axisymmetric drops in a vacuum. \emph{Physics of Fluids}
  \textbf{2018}, \emph{30}, 064107\relax
\mciteBstWouldAddEndPuncttrue
\mciteSetBstMidEndSepPunct{\mcitedefaultmidpunct}
{\mcitedefaultendpunct}{\mcitedefaultseppunct}\relax
\EndOfBibitem
\bibitem[Zhang and Liao(2017)Zhang, and Liao]{zhang_theory_2017}
Zhang,~K.; Liao,~X. \emph{Theory and {Modeling} of {Rotating} {Fluids}:
  {Convection}, {Inertial} {Waves} and {Precession}}; Cambridge University
  Press: Cambridge, 2017\relax
\mciteBstWouldAddEndPuncttrue
\mciteSetBstMidEndSepPunct{\mcitedefaultmidpunct}
{\mcitedefaultendpunct}{\mcitedefaultseppunct}\relax
\EndOfBibitem
\bibitem[Brown and Scriven(1980)Brown, and Scriven]{brown_shape_1980}
Brown,~R.~A.; Scriven,~L.~E. The {Shape} and {Stability} of {Rotating} {Liquid}
  {Drops}. \emph{Proceedings of the Royal Society A: Mathematical, Physical and
  Engineering Sciences} \textbf{1980}, \emph{371}, 331--357\relax
\mciteBstWouldAddEndPuncttrue
\mciteSetBstMidEndSepPunct{\mcitedefaultmidpunct}
{\mcitedefaultendpunct}{\mcitedefaultseppunct}\relax
\EndOfBibitem
\bibitem[Vonnegut(1942)]{vonnegut_rotating_1942}
Vonnegut,~B. Rotating {Bubble} {Method} for the {Determination} of {Surface}
  and {Interfacial} {Tensions}. \emph{Review of Scientific Instruments}
  \textbf{1942}, \emph{13}, 6--9\relax
\mciteBstWouldAddEndPuncttrue
\mciteSetBstMidEndSepPunct{\mcitedefaultmidpunct}
{\mcitedefaultendpunct}{\mcitedefaultseppunct}\relax
\EndOfBibitem
\bibitem[Princen \latin{et~al.}(1967)Princen, Zia, and
  Mason]{princen_measurement_1967}
Princen,~H.; Zia,~I.; Mason,~S. Measurement of interfacial tension from the
  shape of a rotating drop. \emph{Journal of Colloid and Interface Science}
  \textbf{1967}, \emph{23}, 99--107\relax
\mciteBstWouldAddEndPuncttrue
\mciteSetBstMidEndSepPunct{\mcitedefaultmidpunct}
{\mcitedefaultendpunct}{\mcitedefaultseppunct}\relax
\EndOfBibitem
\bibitem[Torza(1975)]{torza_rotatingbubble_1975}
Torza,~S. The rotating-bubble apparatus. \emph{Review of Scientific
  Instruments} \textbf{1975}, \emph{46}, 778--783\relax
\mciteBstWouldAddEndPuncttrue
\mciteSetBstMidEndSepPunct{\mcitedefaultmidpunct}
{\mcitedefaultendpunct}{\mcitedefaultseppunct}\relax
\EndOfBibitem
\bibitem[Cayas \latin{et~al.}()Cayas, Schechter, and Wade]{cayas_acs_nodate}
Cayas,~J.; Schechter,~R.; Wade,~W. \emph{({ACS} {Symposium} {Volume} 8) {K}.
  {L}. {Mittal} ({Eds}.) - {Adsorption} at {Interfaces}-{American} {Chemical}
  {Society} (1975)}\relax
\mciteBstWouldAddEndPuncttrue
\mciteSetBstMidEndSepPunct{\mcitedefaultmidpunct}
{\mcitedefaultendpunct}{\mcitedefaultseppunct}\relax
\EndOfBibitem
\bibitem[Patterson \latin{et~al.}(1971)Patterson, Hu, and
  Grindstaff]{patterson_measurement_1971}
Patterson,~H.~T.; Hu,~K.~H.; Grindstaff,~T.~H. Measurement of interfacial and
  surface tensions in polymer systems. \emph{Journal of Polymer Science Part C:
  Polymer Symposia} \textbf{1971}, \emph{34}, 31--43\relax
\mciteBstWouldAddEndPuncttrue
\mciteSetBstMidEndSepPunct{\mcitedefaultmidpunct}
{\mcitedefaultendpunct}{\mcitedefaultseppunct}\relax
\EndOfBibitem
\bibitem[Hsu and Flumerfelt(1975)Hsu, and Flumerfelt]{hsu_rheological_1975}
Hsu,~J.~C.; Flumerfelt,~R.~W. Rheological {Applications} of a {Drop}
  {Elongation} {Experiment}. \emph{Transactions of the Society of Rheology}
  \textbf{1975}, \emph{19}, 523--540\relax
\mciteBstWouldAddEndPuncttrue
\mciteSetBstMidEndSepPunct{\mcitedefaultmidpunct}
{\mcitedefaultendpunct}{\mcitedefaultseppunct}\relax
\EndOfBibitem
\bibitem[Joseph \latin{et~al.}(1992)Joseph, Arney, Gillberg, Hu, Hultman,
  Verdier, and Vinagre]{joseph_spinning_1992}
Joseph,~D.~D.; Arney,~M.~S.; Gillberg,~G.; Hu,~H.; Hultman,~D.; Verdier,~C.;
  Vinagre,~T.~M. A spinning drop tensioextensometer. \emph{Journal of Rheology}
  \textbf{1992}, \emph{36}, 621--662\relax
\mciteBstWouldAddEndPuncttrue
\mciteSetBstMidEndSepPunct{\mcitedefaultmidpunct}
{\mcitedefaultendpunct}{\mcitedefaultseppunct}\relax
\EndOfBibitem
\bibitem[Joseph \latin{et~al.}(1992)Joseph, Arney, and Ma]{joseph_upper_1992}
Joseph,~D.; Arney,~M.; Ma,~G. Upper and lower bounds for interfacial tension
  using spinning drop devices. \emph{Journal of Colloid and Interface Science}
  \textbf{1992}, \emph{148}, 291--294\relax
\mciteBstWouldAddEndPuncttrue
\mciteSetBstMidEndSepPunct{\mcitedefaultmidpunct}
{\mcitedefaultendpunct}{\mcitedefaultseppunct}\relax
\EndOfBibitem
\bibitem[Hu and Joseph(1994)Hu, and Joseph]{hu_evolution_1994}
Hu,~H.; Joseph,~D. Evolution of a {Liquid} {Drop} in a {Spinning} {Drop}
  {Tensiometer}. \emph{Journal of Colloid and Interface Science} \textbf{1994},
  \emph{162}, 331--339\relax
\mciteBstWouldAddEndPuncttrue
\mciteSetBstMidEndSepPunct{\mcitedefaultmidpunct}
{\mcitedefaultendpunct}{\mcitedefaultseppunct}\relax
\EndOfBibitem
\bibitem[Stone and Bush(1996)Stone, and Bush]{stone_stone_1996}
Stone,~H.; Bush,~J. Stone - {Time}-dep drop deformation in rotating viscous
  fluid.pdf. \emph{Quarterly of Applied Mathematics} \textbf{1996}, \emph{54},
  551--556\relax
\mciteBstWouldAddEndPuncttrue
\mciteSetBstMidEndSepPunct{\mcitedefaultmidpunct}
{\mcitedefaultendpunct}{\mcitedefaultseppunct}\relax
\EndOfBibitem
\bibitem[Than \latin{et~al.}(1988)Than, Preziosi, Joseph, and
  Arney]{than_measurement_1988}
Than,~P.; Preziosi,~L.; Joseph,~D.; Arney,~M. Measurement of {Interfacial}
  {Tension} between {Immiscible} {Liquids} with the {Spinning} {Rod}
  {Tensiometer}. \emph{Journal of Colloid and Interface Science} \textbf{1988},
  \emph{124}, 552--559\relax
\mciteBstWouldAddEndPuncttrue
\mciteSetBstMidEndSepPunct{\mcitedefaultmidpunct}
{\mcitedefaultendpunct}{\mcitedefaultseppunct}\relax
\EndOfBibitem
\bibitem[Verdier(1990)]{verdier_topics_1990}
Verdier,~C. \emph{Topics in the {Fluid} {Dynamics} of {Viscoelastic} {Liquids}.
  {PhD} dissertation}; 1990\relax
\mciteBstWouldAddEndPuncttrue
\mciteSetBstMidEndSepPunct{\mcitedefaultmidpunct}
{\mcitedefaultendpunct}{\mcitedefaultseppunct}\relax
\EndOfBibitem
\bibitem[Manning and Scriven(1977)Manning, and
  Scriven]{manning_interfacial_1977}
Manning,~C.~D.; Scriven,~L.~E. On interfacial tension measurement with a
  spinning drop in gyrostatic equilibrium. \emph{Review of Scientific
  Instruments} \textbf{1977}, \emph{48}, 1699--1705\relax
\mciteBstWouldAddEndPuncttrue
\mciteSetBstMidEndSepPunct{\mcitedefaultmidpunct}
{\mcitedefaultendpunct}{\mcitedefaultseppunct}\relax
\EndOfBibitem
\bibitem[Petitjeans(1996)]{petitjeans_petitjean_1996}
Petitjeans,~P. Petitjean - {Une} tension de surface pour le fluides
  miscibles. \emph{C.R. Acad Sci. Paris} \textbf{1996}, 673--679\relax
\mciteBstWouldAddEndPuncttrue
\mciteSetBstMidEndSepPunct{\mcitedefaultmidpunct}
{\mcitedefaultendpunct}{\mcitedefaultseppunct}\relax
\EndOfBibitem
\bibitem[Pojman \latin{et~al.}(2006)Pojman, Whitmore, Turco~Liveri, Lombardo,
  Marszalek, Parker, and Zoltowski]{pojman_evidence_2006}
Pojman,~J.~A.; Whitmore,~C.; Turco~Liveri,~M.~L.; Lombardo,~R.; Marszalek,~J.;
  Parker,~R.; Zoltowski,~B. Evidence for the {Existence} of an {Effective}
  {Interfacial} {Tension} between {Miscible} {Fluids}: {Isobutyric}
  {Acid}-{Water} and 1-{Butanol}-{Water} in a {Spinning}-{Drop}
  {Tensiometer}. \emph{Langmuir} \textbf{2006}, \emph{22}, 2569--2577\relax
\mciteBstWouldAddEndPuncttrue
\mciteSetBstMidEndSepPunct{\mcitedefaultmidpunct}
{\mcitedefaultendpunct}{\mcitedefaultseppunct}\relax
\EndOfBibitem
\bibitem[Zoltowski \latin{et~al.}(2007)Zoltowski, Chekanov, Masere, Pojman, and
  Volpert]{zoltowski_evidence_2007}
Zoltowski,~B.; Chekanov,~Y.; Masere,~J.; Pojman,~J.~A.; Volpert,~V. Evidence
  for the {Existence} of an {Effective} {Interfacial} {Tension} between
  {Miscible} {Fluids}. 2. {Dodecyl} {Acrylate}-{Poly}(dodecyl acrylate) in a
  {Spinning} {Drop} {Tensiometer}. \emph{Langmuir} \textbf{2007}, \emph{23},
  5522--5531\relax
\mciteBstWouldAddEndPuncttrue
\mciteSetBstMidEndSepPunct{\mcitedefaultmidpunct}
{\mcitedefaultendpunct}{\mcitedefaultseppunct}\relax
\EndOfBibitem
\bibitem[Lister and Stone(1996)Lister, and Stone]{lister_time-dependent_1996}
Lister,~J.; Stone,~H. Time-dependent viscous deformation of a drop in a rapidly
  rotating denser fluid. \emph{Journal of Fluid Mechanics} \textbf{1996},
  \emph{317}, 275--299\relax
\mciteBstWouldAddEndPuncttrue
\mciteSetBstMidEndSepPunct{\mcitedefaultmidpunct}
{\mcitedefaultendpunct}{\mcitedefaultseppunct}\relax
\EndOfBibitem
\bibitem[Saad \latin{et~al.}(2011)Saad, Policova, and
  Neumann]{saad_design_2011}
Saad,~S.~M.; Policova,~Z.; Neumann,~A.~W. Design and accuracy of pendant drop
  methods for surface tension measurement. \emph{Colloids and Surfaces A:
  Physicochemical and Engineering Aspects} \textbf{2011}, \emph{384},
  442--452\relax
\mciteBstWouldAddEndPuncttrue
\mciteSetBstMidEndSepPunct{\mcitedefaultmidpunct}
{\mcitedefaultendpunct}{\mcitedefaultseppunct}\relax
\EndOfBibitem
\bibitem[Rallison(1984)]{rallison_deformation_1984}
Rallison,~J.~M. The {Deformation} of {Small} {Viscous} {Drops} and {Bubbles} in
  {Shear} {Flows}. \emph{Ann. Rev. Fluid Mech.} \textbf{1984}, \emph{16},
  45--66\relax
\mciteBstWouldAddEndPuncttrue
\mciteSetBstMidEndSepPunct{\mcitedefaultmidpunct}
{\mcitedefaultendpunct}{\mcitedefaultseppunct}\relax
\EndOfBibitem
\bibitem[Batchelor(1970)]{batchelor_slender-body_1970}
Batchelor,~G.~K. Slender-body theory for particles of arbitrary cross-section
  in {Stokes} flow. \emph{Journal of Fluid Mechanics} \textbf{1970}, \emph{44},
  419\relax
\mciteBstWouldAddEndPuncttrue
\mciteSetBstMidEndSepPunct{\mcitedefaultmidpunct}
{\mcitedefaultendpunct}{\mcitedefaultseppunct}\relax
\EndOfBibitem
\bibitem[Cox(1970)]{cox_motion_1970}
Cox,~R.~G. The motion of long slender bodies in a viscous fluid {Part} 1.
  {General} theory. \emph{Journal of Fluid Mechanics} \textbf{1970}, \emph{44},
  791\relax
\mciteBstWouldAddEndPuncttrue
\mciteSetBstMidEndSepPunct{\mcitedefaultmidpunct}
{\mcitedefaultendpunct}{\mcitedefaultseppunct}\relax
\EndOfBibitem
\bibitem[noa(1963)]{noauthor_physical_1963}
\emph{Physical properties of glycerine and its solutions}; Glycerine Producers'
  Association: New York, 1963\relax
\mciteBstWouldAddEndPuncttrue
\mciteSetBstMidEndSepPunct{\mcitedefaultmidpunct}
{\mcitedefaultendpunct}{\mcitedefaultseppunct}\relax
\EndOfBibitem
\bibitem[Currie and Van~Nieuwkoop(1982)Currie, and
  Van~Nieuwkoop]{currie_buoyancy_1982}
Currie,~P.; Van~Nieuwkoop,~J. Buoyancy effects in the spinning-drop interfacial
  tensiometer. \emph{Journal of Colloid and Interface Science} \textbf{1982},
  \emph{87}, 301--316\relax
\mciteBstWouldAddEndPuncttrue
\mciteSetBstMidEndSepPunct{\mcitedefaultmidpunct}
{\mcitedefaultendpunct}{\mcitedefaultseppunct}\relax
\EndOfBibitem
\bibitem[Stone \latin{et~al.}(1986)Stone, Bentley, and
  Leal]{stone_experimental_1986}
Stone,~H.~A.; Bentley,~B.~J.; Leal,~L.~G. An experimental study of transient
  effects in the breakup of viscous drops. \emph{Journal of Fluid Mechanics}
  \textbf{1986}, \emph{173}, 131\relax
\mciteBstWouldAddEndPuncttrue
\mciteSetBstMidEndSepPunct{\mcitedefaultmidpunct}
{\mcitedefaultendpunct}{\mcitedefaultseppunct}\relax
\EndOfBibitem
\bibitem[Panchompoo \latin{et~al.}(2012)Panchompoo, Aldous, Baker, Wallace, and
  Compton]{panchompoo_one-step_2012}
Panchompoo,~J.; Aldous,~L.; Baker,~M.; Wallace,~M.~I.; Compton,~R.~G. One-step
  synthesis of fluorescein modified nano-carbon for {Pd}(ii) detection via
  fluorescence quenching. \emph{The Analyst} \textbf{2012}, \emph{137},
  2054\relax
\mciteBstWouldAddEndPuncttrue
\mciteSetBstMidEndSepPunct{\mcitedefaultmidpunct}
{\mcitedefaultendpunct}{\mcitedefaultseppunct}\relax
\EndOfBibitem
\bibitem[Szalay and Tomb\'{a}cz(1964)Szalay, and Tomb\'{a}cz]{szalay_effect_1964}
Szalay,~L.; Tomb\'{a}cz,~E. Effect of the solvent on the fluorescence spectrum of
  trypaflavine and fluorescein. \emph{Acta Physica Academiae Scientiarum
  Hungaricae} \textbf{1964}, \emph{16}, 367--371\relax
\mciteBstWouldAddEndPuncttrue
\mciteSetBstMidEndSepPunct{\mcitedefaultmidpunct}
{\mcitedefaultendpunct}{\mcitedefaultseppunct}\relax
\EndOfBibitem
\bibitem[Viades-Trejo and Gracia-Fadrique(2007)Viades-Trejo, and
  Gracia-Fadrique]{viades-trejo_spinning_2007}
Viades-Trejo,~J.; Gracia-Fadrique,~J. Spinning drop method. \emph{Colloids and
  Surfaces A: Physicochemical and Engineering Aspects} \textbf{2007},
  \emph{302}, 549--552\relax
\mciteBstWouldAddEndPuncttrue
\mciteSetBstMidEndSepPunct{\mcitedefaultmidpunct}
{\mcitedefaultendpunct}{\mcitedefaultseppunct}\relax
\EndOfBibitem
\bibitem[Koos \latin{et~al.}(2012)Koos, Johannsmeier, Schwebler, and
  Willenbacher]{koos_tuning_2012}
Koos,~E.; Johannsmeier,~J.; Schwebler,~L.; Willenbacher,~N. Tuning suspension
  rheology using capillary forces. \emph{Soft Matter} \textbf{2012}, \emph{8},
  6620\relax
\mciteBstWouldAddEndPuncttrue
\mciteSetBstMidEndSepPunct{\mcitedefaultmidpunct}
{\mcitedefaultendpunct}{\mcitedefaultseppunct}\relax
\EndOfBibitem
\bibitem[Sinzato \latin{et~al.}(2015)Sinzato, Dias, and
  Cunha]{sinzato_experimental_2015}
Sinzato,~Y.~Z.; Dias,~N. J.~S.; Cunha,~F.~R. {Experimental} {measurements} {of}
  {interfacial} {tension} {between} {liquid}-{liquid} {mixtures} {in} {the}
  {presence} {of} {surfactants}. Proceedings 23rd {ABCM} {International}
  {Congress} of {Mechanical} {Engineering}. Rio de Janeiro, Brazil, 2015\relax
\mciteBstWouldAddEndPuncttrue
\mciteSetBstMidEndSepPunct{\mcitedefaultmidpunct}
{\mcitedefaultendpunct}{\mcitedefaultseppunct}\relax
\EndOfBibitem
\bibitem[Truzzolillo and Cipelletti(2017)Truzzolillo, and
  Cipelletti]{truzzolillo_off-equilibrium_2017}
Truzzolillo,~D.; Cipelletti,~L. Off-equilibrium surface tension in miscible
  fluids. \emph{Soft Matter} \textbf{2017}, \emph{13}, 13--21\relax
\mciteBstWouldAddEndPuncttrue
\mciteSetBstMidEndSepPunct{\mcitedefaultmidpunct}
{\mcitedefaultendpunct}{\mcitedefaultseppunct}\relax
\EndOfBibitem
\bibitem[Truzzolillo \latin{et~al.}(2016)Truzzolillo, Mora, Dupas, and
  Cipelletti]{truzzolillo_nonequilibrium_2016}
Truzzolillo,~D.; Mora,~S.; Dupas,~C.; Cipelletti,~L. Nonequilibrium
  {Interfacial} {Tension} in {Simple} and {Complex} {Fluids}. \emph{Physical
  Review X} \textbf{2016}, \emph{6}, 041057\relax
\mciteBstWouldAddEndPuncttrue
\mciteSetBstMidEndSepPunct{\mcitedefaultmidpunct}
{\mcitedefaultendpunct}{\mcitedefaultseppunct}\relax
\EndOfBibitem
\bibitem[Obidi \latin{et~al.}(2017)Obidi, Muggeridge, and
  Vesovic]{obidi_analytical_2017}
Obidi,~O.; Muggeridge,~A.~H.; Vesovic,~V. Analytical solution for compositional
  profile driven by gravitational segregation and diffusion. \emph{Physical
  Review E} \textbf{2017}, \emph{95}\relax
\mciteBstWouldAddEndPuncttrue
\mciteSetBstMidEndSepPunct{\mcitedefaultmidpunct}
{\mcitedefaultendpunct}{\mcitedefaultseppunct}\relax
\EndOfBibitem
\bibitem[noa()]{noauthor_supplemental_nodate}
Supplemental {Material}: {Calculation} of the rotational length for
  water-glycerol mixtures.\relax
\mciteBstWouldAddEndPunctfalse
\mciteSetBstMidEndSepPunct{\mcitedefaultmidpunct}
{}{\mcitedefaultseppunct}\relax
\EndOfBibitem
\bibitem[D'Errico \latin{et~al.}(2004)D'Errico, Ortona, Capuano, and
  Vitagliano]{derrico_diffusion_2004}
D'Errico,~G.; Ortona,~O.; Capuano,~F.; Vitagliano,~V. Diffusion {Coefficients}
  for the {Binary} {System} {Glycerol} + {Water} at 25 $^\circ${C}. {A} {Velocity}
  {Correlation} {Study}. \emph{Journal of Chemical \& Engineering Data}
  \textbf{2004}, \emph{49}, 1665--1670\relax
\mciteBstWouldAddEndPuncttrue
\mciteSetBstMidEndSepPunct{\mcitedefaultmidpunct}
{\mcitedefaultendpunct}{\mcitedefaultseppunct}\relax
\EndOfBibitem
\bibitem[Cheng(2008)]{cheng_formula_2008}
Cheng,~N.-S. Formula for the {Viscosity} of a {Glycerol}-{Water} {Mixture}.
  \emph{Industrial \& Engineering Chemistry Research} \textbf{2008}, \emph{47},
  3285--3288\relax
\mciteBstWouldAddEndPuncttrue
\mciteSetBstMidEndSepPunct{\mcitedefaultmidpunct}
{\mcitedefaultendpunct}{\mcitedefaultseppunct}\relax
\EndOfBibitem
\bibitem[Zeldovich(1949)]{Zeldovich1949}
Zeldovich,~Y.~B. \emph{Zh. Fiz. Khim. (in Russian) 1949, 23, 931-935}
  \textbf{1949}, \emph{23}, 931--935\relax
\mciteBstWouldAddEndPuncttrue
\mciteSetBstMidEndSepPunct{\mcitedefaultmidpunct}
{\mcitedefaultendpunct}{\mcitedefaultseppunct}\relax
\EndOfBibitem
\end{mcitethebibliography}

\providecommand{\latin}[1]{#1}
\makeatletter
\providecommand{\doi}
  {\begingroup\let\do\@makeother\dospecials
  \catcode`\{=1 \catcode`\}=2 \doi@aux}
\providecommand{\doi@aux}[1]{\endgroup\texttt{#1}}
\makeatother
\providecommand*\mcitethebibliography{\thebibliography}
\csname @ifundefined\endcsname{endmcitethebibliography}
  {\let\endmcitethebibliography\endthebibliography}{}



\begin{tocentry}

\begin{center}
\includegraphics[width=6cm,height=3.5cm]{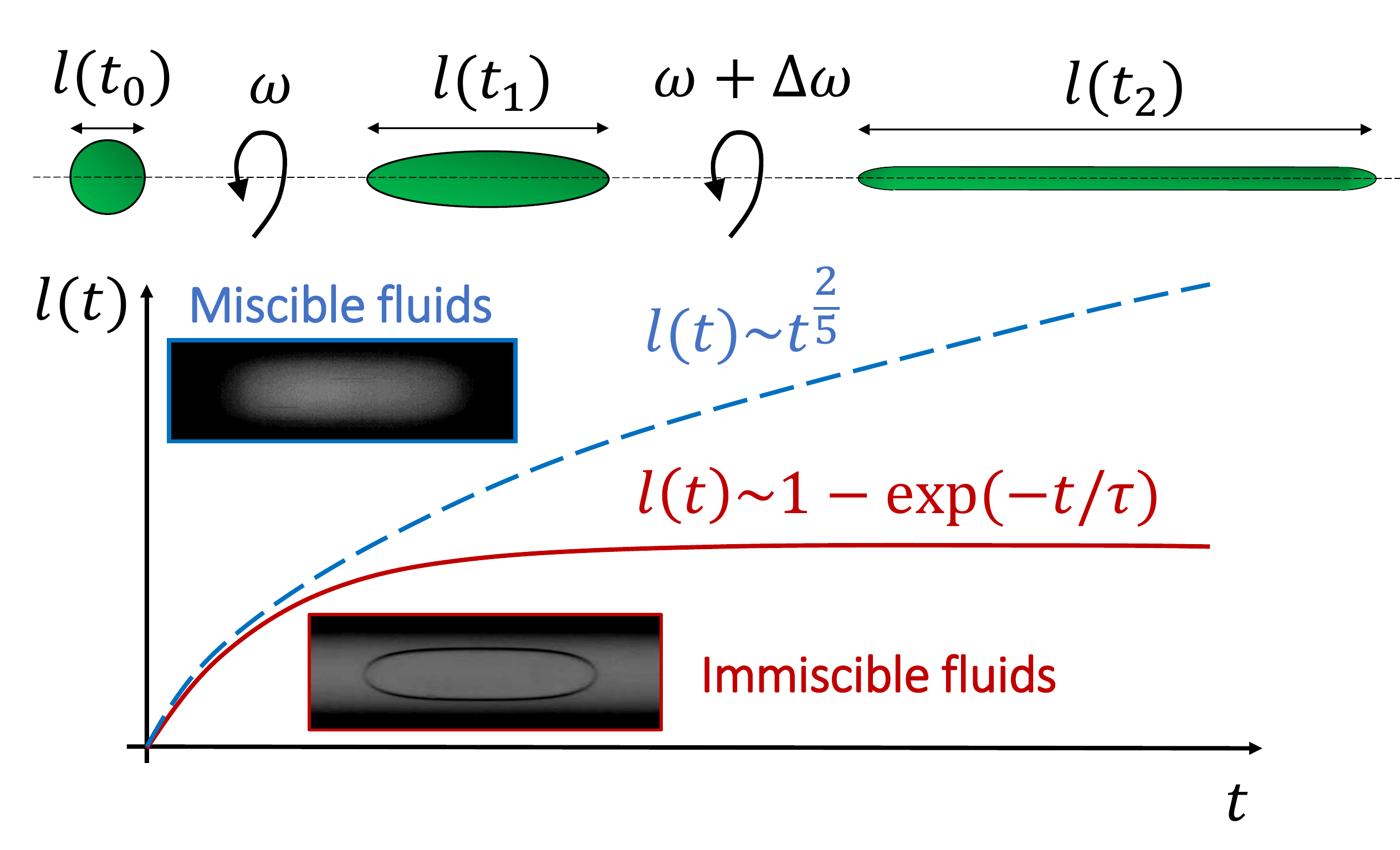}
\end{center}

\end{tocentry}

\end{document}